\documentclass[aps,prd,twocolumn,10pt,superscriptaddress,nofootinbib,floatfix,longbibliography]{revtex4-2}
\usepackage[caption=false]{subfig}
\usepackage{mathrsfs,amsmath,amssymb,graphicx,bm,booktabs,hyperref,orcidlink}
\hypersetup{colorlinks=true,citecolor=blue,linkcolor=blue,urlcolor=blue}
\newcommand{\ii}{\mathrm{i}}

\newcommand{\dd}{\mathrm{d}}

\begin{document}
\title{Complete quasinormal modes of Type-D black holes. III. Reissner--Nordstr\"om spectra from zero to extremal charge}
\author{Changkai Chen\,\orcidlink{0000-0002-4023-0682}}
\affiliation{School of Physics and Astronomy, Beijing Normal University, Beijing, 100875, China}

\author{Xiaohua Zhang\,\orcidlink{0000-0002-0914-5455}}\email[Contact author: ]{zhangxiaohua@ynnu.edu.cn}
\affiliation{School of Mathematics, Yunnan Normal University, Kunming 650500, China}

\author{Zhoujian Cao\,\orcidlink{0000-0002-1932-7295} } \email[Contact author: ]{zjcao@bnu.edu.cn}
\affiliation{School of Physics and Astronomy, Beijing Normal University, Beijing, 100875, China}
\affiliation{Institute for Frontiers in Astronomy and Astrophysics, Beijing Normal University, Beijing, 102206, China}

\affiliation{School of Fundamental Physics and Mathematical Sciences, Hangzhou Institute for Advanced Study, UCAS, Hangzhou, 310024, China}

\author{Jiliang Jing\,\orcidlink{0000-0002-2803-7900} }
\affiliation{Department of Physics, Key Laboratory of Low Dimensional Quantum Structures and Quantum Control of Ministry of Education, Synergetic Innovation Center for Quantum Effects and Applications, Hunan Normal University, Changsha, 410081, Hunan, China}
\date{\today}
\begin{abstract}
Earlier calculations left open how gravitational quasinormal modes (QNMs) of charged black holes recover the Schwarzschild spectrum and how the modes near algebraically special frequencies relate to the exact special solutions. We extend our Kerr analysis to Reissner--Nordstr\"om black holes by analytically continuing the ingoing horizon solution and locating the zeros of the incoming amplitude at infinity. For each of $\ell=2$ and $3$, we trace the first 42 overtone families from the Schwarzschild limit to extremality. The charge trajectories distinguish crossings of the negative imaginary axis from turns near it and separate the unconventional Schwarzschild modes from neighboring branches. At zero charge the polar algebraically special mode is a QNM. At extremal charge the exact axial special modes are total-transmission modes. Of the 84 families, 75 approach distinct extremal frequencies with finite damping, and nine approach zero frequency with vanishing excitation. At exact extremality we also compute, for the first time, gravitational QNMs on the unconventional Riemann sheet reached across the negative imaginary axis. At intermediate charge, the clearest resonant enhancement of both excitation factors occurs for the $\ell=3$ overtone pairs $(22,23)$ and $(23,24)$.
\end{abstract}
\maketitle
\raggedbottom
\section{Introduction}\label{sec:intro}

Quasinormal modes (QNMs) describe the damped response of a black hole to perturbations. Their frequencies are fixed by the spacetime and the wave conditions at the horizon and infinity \cite{Dreyer:2003bv,Berti:2005ys,Berti:2009kk,Berti:2025hly}. Simulated ringdown waveforms can be used to train models that infer the remnant mass and spin from gravitational-wave data \cite{wanghe,Mishra:2025vjp,Liu:2026nhu}. To understand how a spectrum changes with charge or spin, one must also identify the individual mode families, their Schwarzschild origins, and their extremal limits \cite{Berti:2004md}.

Berti posed two questions that motivate this survey \cite{Berti:2004md}. How do the spectra of rotating or charged black holes recover the Schwarzschild spectrum as spin or charge vanishes? How are the QNMs near an algebraically special (AS) frequency related to the exact special solution \cite{Berti:2009kk}? Both questions require tracking individual families. For asymptotically large damping the order of the zero-charge and infinite-damping limits also matters \cite{Berti:2003zu,Motl:2003cd}.

Spherical symmetry reduces the coupled gravitational and electromagnetic perturbations of Reissner--Nordstr\"om (RN) black holes to radial master equations with known angular eigenvalues. Continued fractions give the nonextremal frequencies \cite{Leaver:1990zz}. Early charge trajectories resolved the spirals of higher overtones but left open whether gravitational branches cross the negative imaginary axis \cite{Andersson:1996xw}. Berti and Kokkotas traced oscillations with charge up to overtone numbers of order $10^5$ \cite{Berti:2003zu}. For the first and second gravitational overtones at $\ell=2$, Berti and Cardoso located the first extrema of $\Re(M\omega)$ at $Q/M=0.958$ and $0.910$, respectively \cite{Berti:2008xu}.

In asymptotically flat spacetimes the Green function contains isolated poles and a branch contribution associated with the negative imaginary axis. Tracking a mode through that axis requires continuing its boundary solutions across the frequency cut. Unconventional Schwarzschild resonances on the neighboring sheet near the AS frequency provide additional possible parents for charged or rotating families \cite{Leung:2003eq}. At the exact AS frequency the transformation between axial and polar variables is singular, so the radiation conditions must be checked separately in each parity \cite{MaassenvandenBrink:2000iwh}.

At $Q=M$ the horizons merge, the surface gravity vanishes, and the horizon becomes an irregular singular point. Direct extremal calculations give finite-damping roots \cite{Onozawa:1995vu,Daghigh:2024wcl}, including low-lying frequencies accessible to bound-state continuation and complex scaling \cite{Zhou:2026kgp,Ogawa:2026veu}. Near-extremal RN also supports gravitational zero-damping modes \cite{Zimmerman:2015trm}. Their near-horizon region can be matched to the exterior \cite{Porfyriadis:2018yag,Porfyriadis:2018jlw}. The branch contribution to the response has been studied in hyperboloidal and scattering formulations \cite{PanossoMacedo:2018hab,Gajic:2019qdd}.

The spirals in the RN frequency plane raise a further question about the response \cite{Berti:2004md}. Two trajectories can intersect at different charges. We identify resonant enhancement by comparing both excitation factors as the two frequencies approach one another at the same $Q$.

Our previous work addressed the Kerr manifestation of these spectral difficulties \cite{Chen:2025sbz}. Analytic continuation recovered the portions omitted when modes crossed the negative imaginary axis and retained the unconventional Schwarzschild parents. This restored the family connections underlying the apparent loss of modes and the anomalous multiplicities near AS frequencies. That work identified RN as the next application of the same principle: the radial boundary solutions must be continued even where a conventional frequency representation stops. Related studies of Kerr total transmission modes and strict extremal QNMs further distinguish zeros of different scattering amplitudes and the endpoints of the continued families \cite{Chen:2026ane,Chen:2026extremal}.

The same boundary solutions enter radiation calculations with a source. Heun representations and their connection coefficients have been used for gravitational, electromagnetic, and scalar fluxes \cite{Chen:2023ese}, including generic orbits with high eccentricity in near extremal Kerr spacetime \cite{Chen:2023lsa}. Adaptive analytic continuation combined with source integration extends these calculations over orbital parameters \cite{Chen:2026bym}. These applications supply methods for evaluating the scattering amplitudes needed to identify and excite the modes.

We trace the gravitational QNM spectrum of RN black holes from the Schwarzschild limit to extremality, with particular attention to the mode families near the algebraically special frequencies. We also examine how the excitation of two modes changes as their frequencies approach one another.

We begin in Sec.~\ref{sec:radial} with the radial equation and the boundary conditions at the horizon and at infinity. With these conventions in place, Sec.~\ref{sec:method} develops the analytic continuation that determines the incoming amplitude. In Sec.~\ref{sec:results} we present the gravitational QNM spectrum from zero to extremal charge and use excitation factors to identify resonances between overtones. Section~\ref{sec:conclusions} summarizes our findings. Appendix~\ref{app:series} derives the radial series and specifies their continuation. Appendix~\ref{app:frequencies} lists the endpoint frequencies and gives detailed views of the trajectories. Throughout we set $G=c=M=1$ and take $0\leq Q\leq1$. With time dependence ${\rm{e}}^{-\ii\omega t}$, damped modes have $\Im(\omega)<0$.

\section{Radial problem and boundary conventions}\label{sec:radial}

We study the gravitational branch of the coupled electromagnetic and gravitational perturbations at $\ell=2,3$. Its axial potential approaches the Regge--Wheeler potential as $Q\to0$.

\begin{equation}
 \frac{\dd^2R_\ell}{\dd r_*^2}+[\omega^2-V(r)]R_\ell=0,
 \qquad \frac{\dd r_*}{\dd r}=\frac1f ,
 \label{eq:master}
\end{equation}

\begin{subequations}\label{eq:potential}
\begin{align}
 f(r)&=1-\frac2r+\frac{Q^2}{r^2},\qquad r_\pm=1\pm\sqrt{1-Q^2},\label{eq:geometry}\\
 V(r)&=f(r)\left(\frac{L}{r^2}-\frac{q_+}{r^3}
                    +\frac{4Q^2}{r^4}\right),\quad L=\ell(\ell+1),\\
 q_+&=3+\sqrt{9+4Q^2(\ell-1)(\ell+2)} .
\end{align}
\end{subequations}
For asymptotically flat RN black holes in Einstein--Maxwell theory, the axial and polar equations have different effective potentials but are related by the Chandrasekhar transformation~\cite{Chandrasekhar1979RN}. Where the transformation is invertible, it maps an axial QNM to a polar QNM at the same frequency for fixed $\ell$ and $Q$ on the same frequency sheet \cite{Andersson:1996xw}. We use the axial equation to find these frequencies. At algebraically special frequencies the transformation can degenerate, so the radiation conditions must be checked separately in each parity \cite{MaassenvandenBrink:2000iwh}.

A QNM is purely ingoing at the outer horizon and purely outgoing at infinity. Its radial solution has the asymptotic behavior
\begin{equation}
 R_\ell(r)\propto
 \begin{cases}
 {\rm{e}}^{-\ii\omega r_*}, & r\to r_+,\\
 {\rm{e}}^{+\ii\omega r_*}, & |r|\to\infty .
 \end{cases}
 \label{eq:qnm-radiation}
\end{equation}
Logarithms in the tortoise coordinate are defined on the real exterior and continued along the radial path. The same path homotopy and continuous arguments are retained as $\Re(\omega)$ changes sign.

For $Q<1$ we use

\begin{equation}
 r_*=r+\frac{r_+^2}{r_+-r_-}\log(r-r_+)
          -\frac{r_-^2}{r_+-r_-}\log(r-r_-) .
 \label{eq:tortoise}
\end{equation}
Following the notation of our previous work \cite{Chen:2025sbz}, each overtone family retains its label $\hat n$ as $Q$ varies, with $\hat n=0$ for the fundamental mode. The AS frequency of this branch is \cite[Eq.~(4.2)]{Daghigh:2024wcl}

\begin{equation}
 \omega_{\rm AS}(Q)=-\ii\,\frac{L(L-2)}{2q_+}.
 \label{eq:as}
\end{equation}

At zero charge and exact extremal charge, respectively, this gives

\begin{subequations}\label{eq:aslimits}
\begin{align}
 \omega_{\rm AS}(Q=0)&=-\frac{\ii}{12}(\ell-1)\ell(\ell+1)(\ell+2),\\
 \omega_{\rm AS}(Q=M)&=-\frac{\ii}{4}(\ell-1)\ell(\ell+1).
\end{align}
\end{subequations}
For $\ell=2,3$, the Schwarzschild AS frequencies are $-2\ii$ and $-10\ii$ and admit polar QNMs \cite{MaassenvandenBrink:2000iwh}. The corresponding extremal RN frequencies are $-3\ii/2$ and $-6\ii$. The exact axial solutions there have zero reflected amplitude and nonzero incoming amplitude (Eq.~\eqref{eq:AScontrol}), which identifies them as total-transmission modes.

\section{Analytic continuation of the radial solution}\label{sec:method}

\subsection{Convergent subextremal horizon solution}

For $Q<1$, introduce $x=(r-r_+)/(r-r_-)$ and factor the exact ingoing phase as $R_{\ell}^{\mathrm{in}}={\rm{e}}^{-\ii\omega r_*}h(x)$. The normalized remainder has $h(0)=1$; below, $h(r)$ abbreviates $h[x(r)]$ when a radial argument is used. Factoring the phase before generating coefficients preserves the regular horizon variable directly and avoids reconstructing it from cancellation between separately evaluated singular factors.

\begin{subequations}\label{eq:coordinates}
\begin{align}
 r&=\frac{r_+-r_-x}{1-x},\\
 h(x)&=\sum_{n=0}^\infty a_nx^n,\qquad a_0=1 .
\end{align}
\end{subequations}

The series supplies purely ingoing local data within $|x|<1$, with coefficients determined by the radial equation. The normalization is $a_0=1$, and the outer-horizon surface gravity $\kappa_+=(r_+-r_-)/(2r_+^2)$ sets the horizon-index scale. The recurrence and its resonant construction are given in Appendix~\ref{app:series}.

When the horizon power series becomes poorly conditioned, local monodromy provides another way to select the ingoing solution. Let $\mathbf y=(h,\dd h/\dd x)^{\mathsf T}$ at $x_0\in(0,1)$, and let $\mathsf M$ propagate this state counterclockwise around $|x|=x_0$. The regular solution has monodromy eigenvalue 1. The second local solution has exponent $\eta_{\rm H}=\ii\omega/\kappa_+$ and eigenvalue $\mu_{\rm H}={\rm{e}}^{2\pi\ii\eta_{\rm H}}$. Away from integer $\eta_{\rm H}$, its contribution is removed by the projector
\begin{equation}
 \mathsf P_{\rm in}=\frac{\mathsf M-\mu_{\rm H}\mathsf I}{1-\mu_{\rm H}},
 \qquad \mathbf y_{\rm in}=\mathsf P_{\rm in}\mathbf v,
 \qquad \mu_{\rm H}\ne1 .
 \label{eq:horizonprojector}
\end{equation}
We use $\mathbf v=(1,0)^{\mathsf T}$ or $(0,1)^{\mathsf T}$ and normalize the nonzero projected solution to $h(r_0)=1$. Local Taylor expansions evaluate the loop and then propagate the selected solution along the usual exterior contour. The auxiliary loop constructs the horizon data; it does not add a winding to the boundary continuation used for the spectrum. This normalization preserves the frequency zeros but differs from $h(0)=1$, so these amplitudes are used only to locate frequencies. The excitation factors below retain their separately specified axial normalization. At $\mu_{\rm H}=1$, Eq.~\eqref{eq:horizonprojector} is singular and the resonant construction must be treated separately.

\subsection{Infinity sectors and the incoming-amplitude constraint}

The two infinity branches are defined by their radiation factors and asymptotic remainders:

\begin{subequations}\label{eq:directbranches}
\begin{align}
 R^{\infty}_{\mathrm{in}}&={\rm{e}}^{-\ii\omega r_*}g_{\infty}^{\mathrm{in}}(1/r),\\
 R^{\infty}_{\mathrm{up}}&={\rm{e}}^{+\ii\omega r_*}g_{\infty}^{\mathrm{up}}(1/r),\\
 g_{\infty}^{\mathrm{in}}&\sim1+\frac{L}{2\ii\omega r}+O(r^{-2}),\\
 g_{\infty}^{\mathrm{up}}&\sim1-\frac{L}{2\ii\omega r}+O(r^{-2}).
\end{align}
\end{subequations}

Choose a ray $r=\rho {\rm{e}}^{\ii\theta}$ satisfying $\Im(\omega {\rm{e}}^{\ii\theta})>0$. The outgoing exponential is then subdominant, so the connection formula gives the incoming amplitude as the ratio $B^{\rm inc}=\lim_{\rho\to\infty}(h/g_{\infty}^{\mathrm{in}})$, evaluated along this ray. The QNM frequencies satisfy $B^{\rm inc}(\omega_{\hat n},Q)=0$, giving the outgoing condition at infinity in Eq.~\eqref{eq:qnm-radiation}. The exponential-suppression test retains the logarithmic terms in the full tortoise coordinate.

\begin{equation}
 \theta=\frac{\pi}{2}(1+\epsilon_\theta)-\arg_{\rm c}\omega+2k\pi,
 \quad |\epsilon_\theta|<1 .
 \label{eq:directangle}
\end{equation}

\begin{equation}
 \frac{R^{\infty}_{\mathrm{up}}}{R^{\infty}_{\mathrm{in}}}
 ={\rm{e}}^{2\ii\omega r_*}\frac{g_{\infty}^{\mathrm{up}}}{g_{\infty}^{\mathrm{in}}}
 \longrightarrow0 .
 \label{eq:directsuppression}
\end{equation}

\begin{equation}
 R_{\ell}^{\mathrm{in}}=B^{\rm inc}R^{\infty}_{\mathrm{in}}+B^{\rm ref}R^{\infty}_{\mathrm{up}} .
 \label{eq:connection}
\end{equation}

\begin{subequations}\label{eq:directbinc}
\begin{align}
 B^{\rm inc}(\omega,Q)
 &=\lim_{\rho\to\infty}\frac{R_{\ell}^{\mathrm{in}}(\rho {\rm{e}}^{\ii\theta})}
                                {R^{\infty}_{\mathrm{in}}(\rho {\rm{e}}^{\ii\theta})}\nonumber\\
 &=\lim_{\rho\to\infty}\frac{h(\rho {\rm{e}}^{\ii\theta})}{g_{\infty}^{\mathrm{in}}({\rm{e}}^{-\ii\theta}/\rho)},\label{eq:binclimit}\\
 B_\rho(\omega,Q)&=\frac{h(\rho {\rm{e}}^{\ii\theta})}
                         {g_{\infty}^{\mathrm{in}}({\rm{e}}^{-\ii\theta}/\rho)} .\label{eq:finitebinc}
\end{align}
\end{subequations}
The integer $k$ and continuous argument in Eq.~\eqref{eq:directangle} specify the continuously followed ray on which the outgoing contribution decays.

\begin{table*}[!t]
\centering\small
\caption{Frequencies on either side of the imaginary-axis crossing for each of the eight $\ell=3$ overtone families. The subscripts $+$ and $-$ denote positive and negative $\Re(\omega)$, respectively, in order of increasing $Q$.}\label{tab:direct80crossings}
\begin{tabular*}{\textwidth}{@{\extracolsep{\fill}}crrr|rrr@{}}\toprule
$\hat{n}$&$Q_+$&$\Re(\omega_+)$&$\Im(\omega_+)$&$Q_-$&$\Re(\omega_-)$&$\Im(\omega_-)$\\\midrule
32 & 0.542300 & 0.000308684 & -7.939640493 & 0.542400 & -0.000142038 & -7.939615676 \\
33 & 0.488000 & 0.001917113 & -8.210578945 & 0.489000 & -0.001839506 & -8.212972580 \\
34 & 0.436400 & 0.000929127 & -8.474781370 & 0.437073 & -0.000314387 & -8.476736405 \\
35 & 0.386125 & 0.000005006 & -8.735742361 & 0.386625 & -0.000338906 & -8.737009064 \\
36 & 0.334454 & 0.000156668 & -8.991087642 & 0.335127 & -0.000046823 & -8.992371299 \\
37 & 0.281233 & 0.000057014 & -9.245326972 & 0.281906 & -0.000015696 & -9.246297383 \\
38 & 0.223183 & 0.000017371 & -9.497911247 & 0.223856 & -0.000003988 & -9.498617823 \\
39 & 0.153460 & 0.000003496 & -9.749311402 & 0.154133 & -0.000000411 & -9.749766549 \\
\bottomrule\end{tabular*}
\end{table*}

\subsection{Radial transport and the extremal limit}

The horizon state $(h,\partial_rh)$ is propagated through overlapping ordinary-point Taylor disks, with rescalings restored to retain absolute normalization. The radial path varies continuously within the same homotopy class as the frequency crosses the imaginary axis. The propagation equation and radial paths are given in Appendix~\ref{app:transport}.

At extremal charge $Q=M$, the horizons merge. The horizon becomes an irregular singular point, so the subextremal Frobenius recurrence is replaced by a separate asymptotic construction. This change also underlies the ordinary-point continued fraction treatment of Onozawa et al.~\cite{Onozawa:1995vu} and its high damping extension \cite{Daghigh:2024wcl}. The radiation conditions and single ended amplitude constraint are retained.

\begin{equation}
 r_*=r-\frac{1}{r-1}+2\log(r-1).
 \label{eq:xrtortoise}
\end{equation}

Here we still write $R_{\ell}^{\mathrm{in}}={\rm{e}}^{-\ii\omega r_*}h$, but initialize $h$ by a normalized asymptotic series in $x=(r-1)/r$, rather than substituting $Q=1$ into the subextremal power series. In the sector $\Im(\omega/x)>0$, we truncate the series at the minimum of its term envelope and continue the solution along the path in Appendix~\ref{app:extremalconstruction} to locate the zeros of the incoming amplitude in Eq.~\eqref{eq:binclimit}. Varying the horizon starting point, expansion order, and extraction radius changed each of the two extremal fundamental frequencies ($\ell=2,3$) by less than $2\times10^{-38}$.

\section{Numerical results}\label{sec:results}\label{sec:numerics}

We locate the zeros of the incoming amplitude in Eq.~\eqref{eq:binclimit} and check the residual $B_\rho$ in Eq.~\eqref{eq:finitebinc} at two extraction radii.

\subsection{RN quasinormal spectrum}

\begin{figure*}[!tp]
\centering
\includegraphics[width=\textwidth,height=.893\textheight,keepaspectratio]{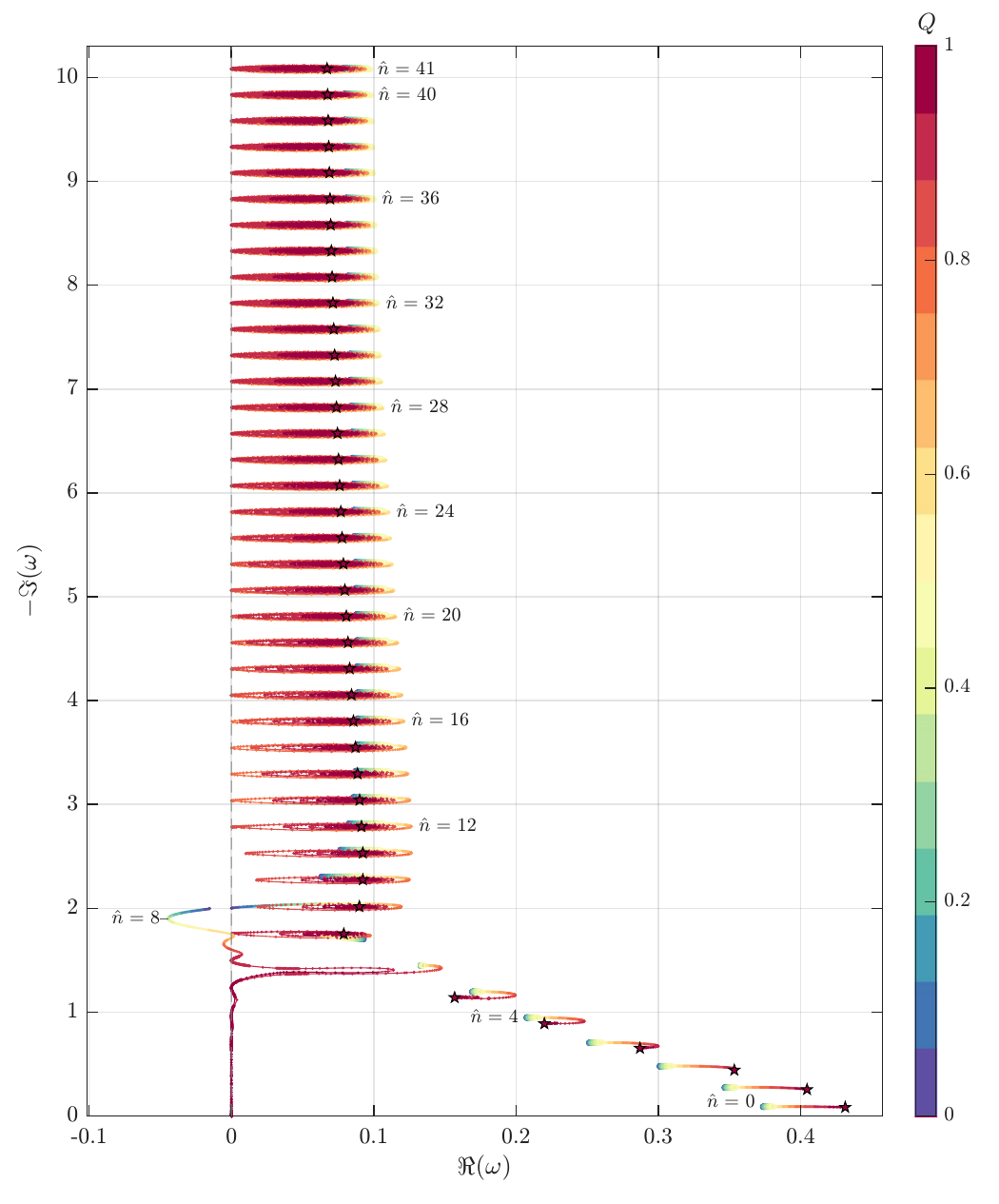}
\caption{Gravitational RN spectrum, $\ell=2$, $\hat n=0,\ldots,41$. Filled circles are direct solutions, open circles are low-charge continued-fraction comparisons, and stars mark $Q=1$. Lines connect successive samples, colored by $Q$.}
\label{fig:summary42l2}
\end{figure*}

\begin{figure*}[!tp]
\centering
\includegraphics[width=\textwidth,height=.893\textheight,keepaspectratio]{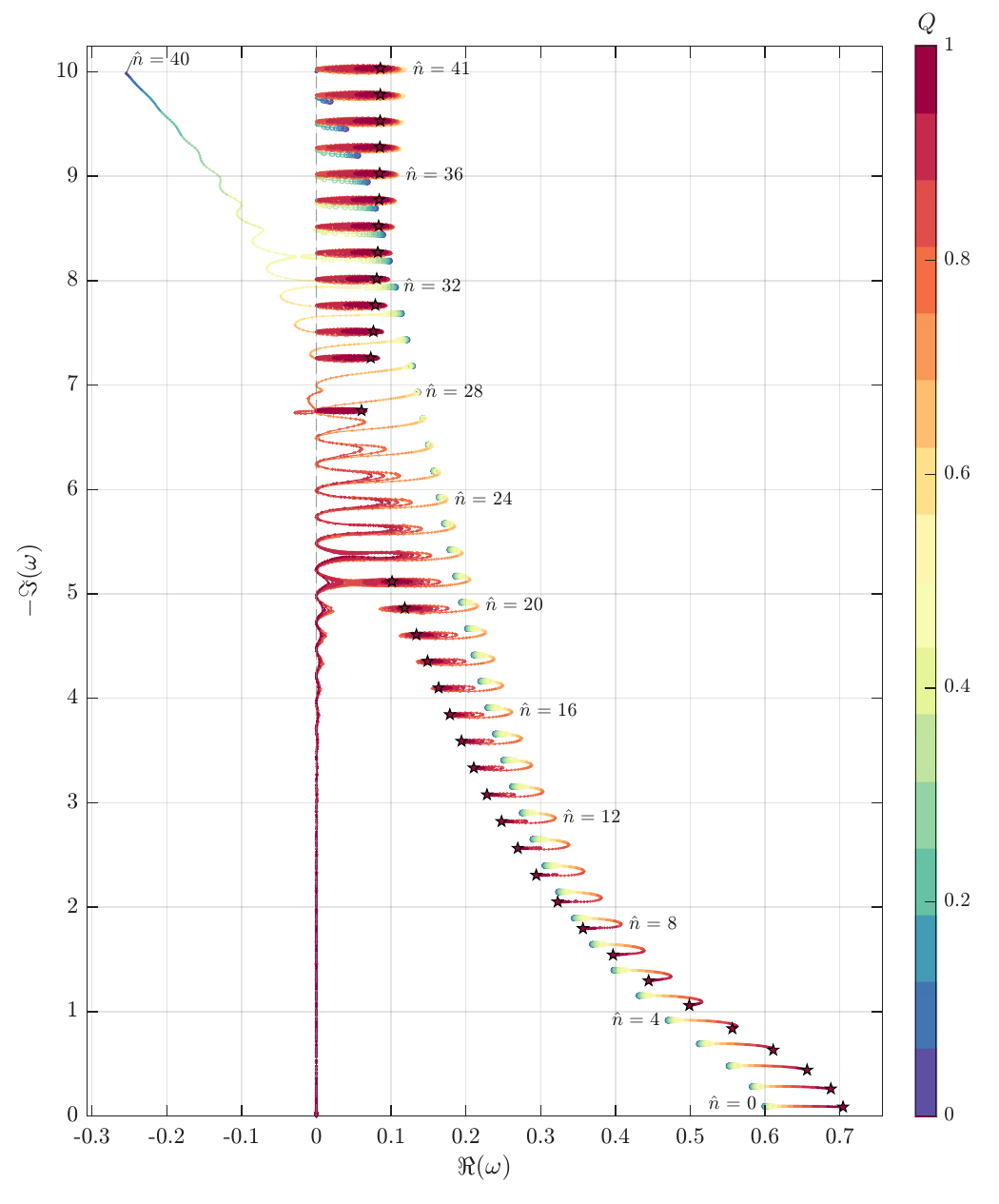}
\caption{Gravitational RN spectrum for $\ell=3$ and $\hat n=0,\ldots,41$. Eight consecutive overtones ($\hat n=32$--39) cross the negative imaginary axis. Symbols and colors follow Fig.~\ref{fig:summary42l2}.}
\label{fig:summary42l3}
\end{figure*}

Figures~\ref{fig:summary42l2} and~\ref{fig:summary42l3} show how the RN gravitational frequencies of the first forty-two overtone families evolve with charge for $\ell=2$ and $3$, respectively. The sign of $\Re(\omega)$ along a trajectory distinguishes a crossing of the imaginary axis from a close turn. At $\ell=3$, eight consecutive overtones, $\hat n=32,\ldots,39$, cross into the negative-real-frequency half-plane. Table~\ref{tab:direct80crossings} gives the frequencies on either side of each crossing. The crossing shifts to lower charge with increasing overtone, from $Q\simeq0.542$ for $\hat n=32$ to $Q\simeq0.154$ for $\hat n=39$. At $\ell=2$, $\hat n=6$ and $17$ turn on the positive side. Refining the charge steps near these turns at $Q\simeq0.9140$ and $0.8540$ gives minimum sampled real parts of approximately $3.32\times10^{-10}$ and $5.76\times10^{-8}$, respectively.

\begin{figure*}[!htbp]\centering
\subfloat[$\ell=2$, $\hat n=8$]{\includegraphics[width=.498\textwidth]{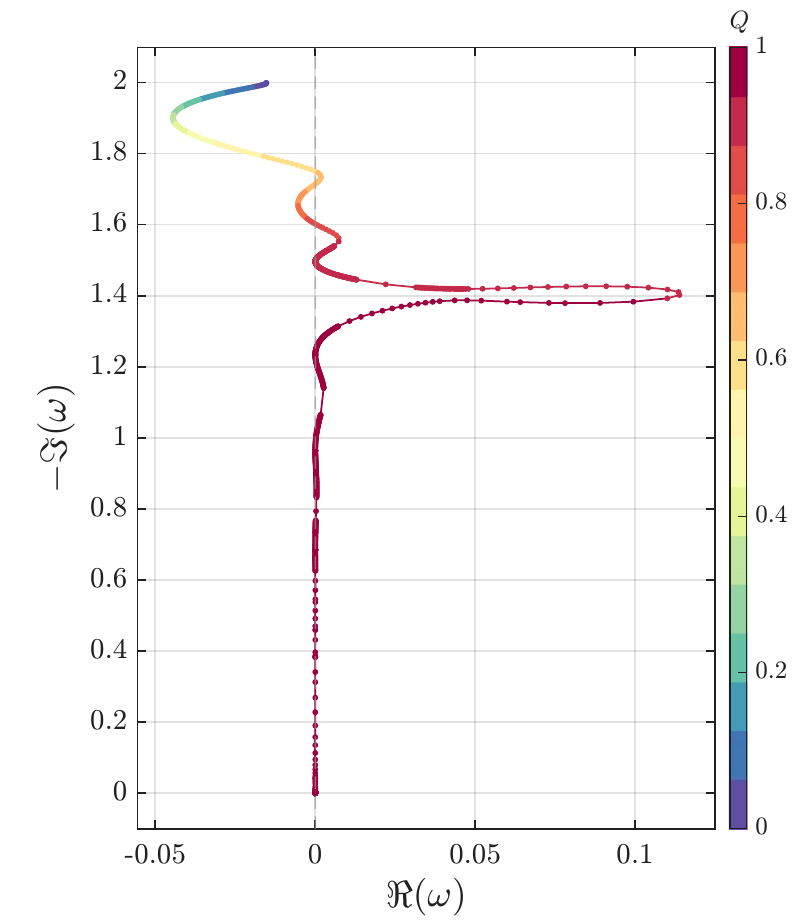}}\hfill
\subfloat[$\ell=2$, $\hat n=9$]{\includegraphics[width=.498\textwidth]{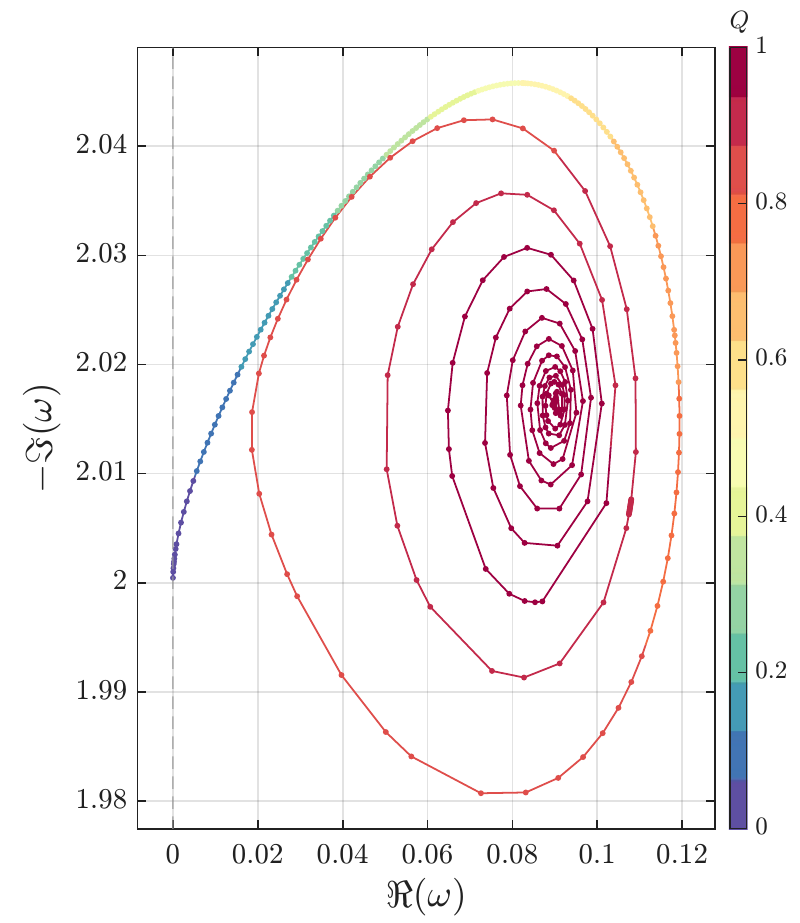}}\par\smallskip
\subfloat[$\ell=3$, $\hat n=40$]{\includegraphics[width=.498\textwidth]{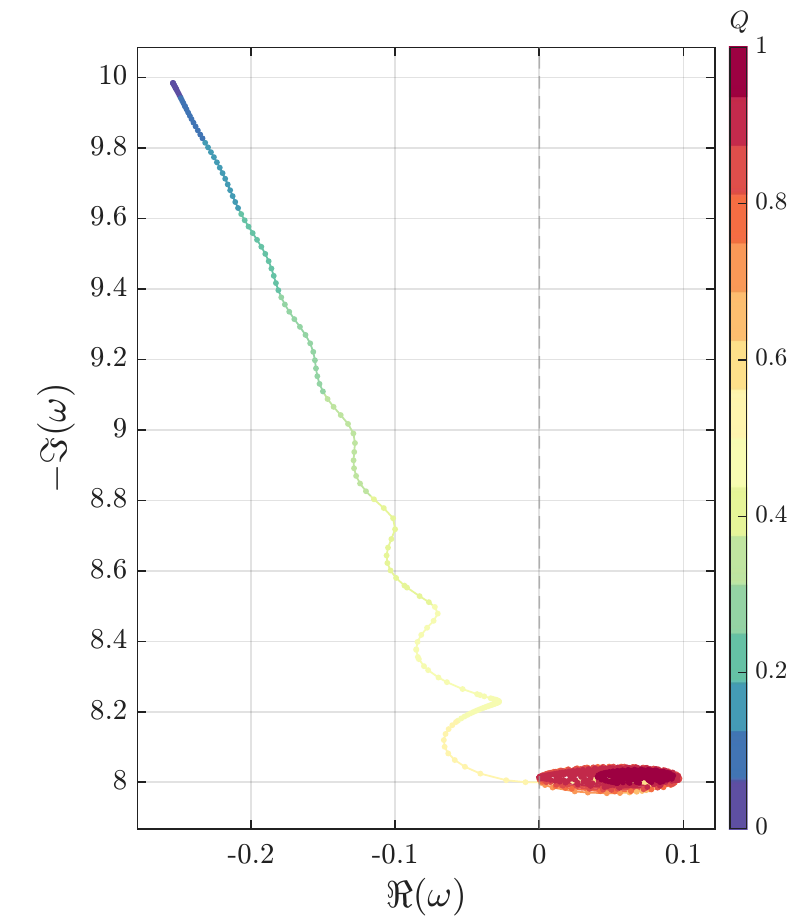}}\hfill
\subfloat[$\ell=3$, $\hat n=41$]{\includegraphics[width=.498\textwidth]{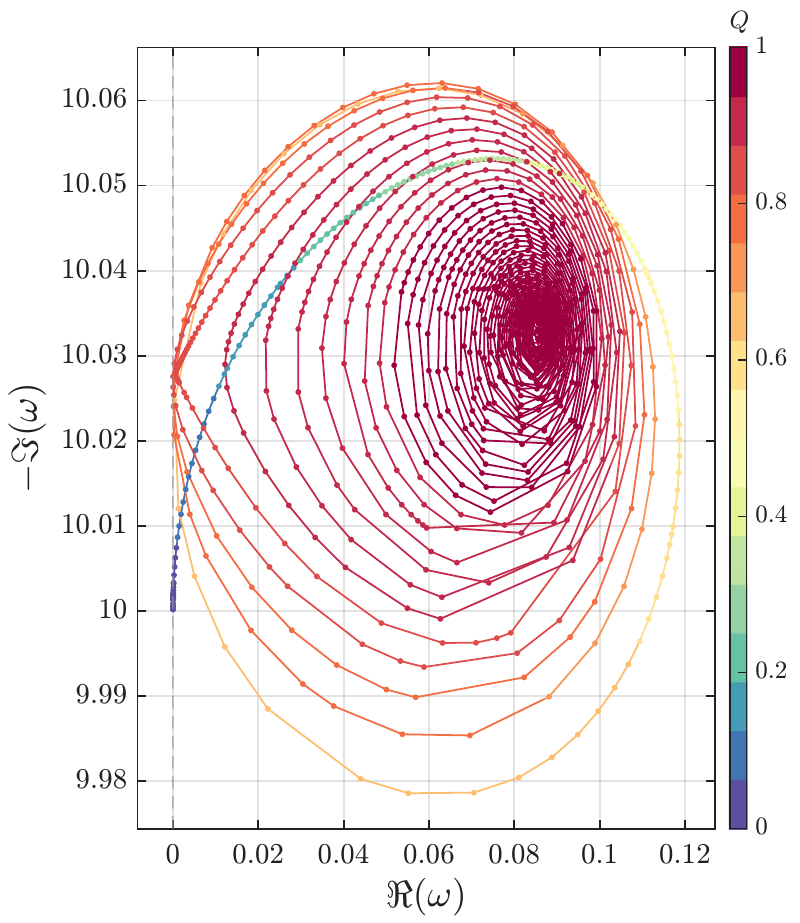}}
\caption{Charge evolution of the unconventional families (left) and AS-neighborhood families (right). Of these four, only the $\ell=2$, $\hat n=8$ family approaches zero frequency.}
\label{fig:selectedfamilies}\end{figure*}

Andersson and Onozawa found an $\ell=2$ gravitational branch approaching $M\omega=-2\ii$ as $Q\to0$ \cite{Andersson:1996xw}. Figure~\ref{fig:selectedfamilies} resolves two families near this frequency and shows that they approach different extremal limits. The unconventional Schwarzschild mode $\hat n=8$ crosses the imaginary axis and approaches zero frequency. The adjacent AS-neighborhood branch $\hat n=9$ spirals toward a finite damping rate $-\Im(\omega)\simeq2.017$. Near the $\ell=3$ AS frequency $M\omega=-10\ii$, both the unconventional branch $\hat n=40$ and the AS-neighborhood branch $\hat n=41$ retain finite damping, with damping rates of $8.019$ and $10.033$, respectively, at $Q=0.99999$.

At the Schwarzschild endpoint ($a=Q=0$), the polar Zerilli solution at the AS frequency is both a QNM and a total-transmission mode. The axial Regge--Wheeler equation has no QNM at that frequency \cite{MaassenvandenBrink:2000iwh,Chen:2025sbz}. At the other endpoint ($Q=1$), the exact axial RN solution is $R_{\rm AS}={\rm{e}}^{-\ii\omega_{\rm AS}r_*}h_{\rm AS}$, with

\begin{subequations}\label{eq:AScontrol}
\begin{align}
 C_{\rm AS}&=\frac{q_+}{L-2},\qquad
 h_{\rm AS}(r)=\frac{1+C_{\rm AS}/r}{1+C_{\rm AS}/r_+},\\
 B^{\rm inc}_{\rm AS}&=\frac{1}{1+C_{\rm AS}/r_+}
 =\begin{cases}1/3,&Q=1,\ \ell=2,\\1/2,&Q=1,\ \ell=3.\end{cases}
\end{align}
\end{subequations}

At $Q=1$ the exact axial AS solutions contain only the ingoing wave at infinity, so $B^{\rm ref}_{\rm AS}=0$. Equation~\eqref{eq:AScontrol} gives incoming amplitudes of $1/3$ for $\ell=2$ and $1/2$ for $\ell=3$. By the scattering-amplitude criterion used in our Kerr study \cite{Chen:2026ane}, both solutions are total-transmission modes for incidence from infinity. Their nonzero incoming amplitudes exclude them from the QNM spectrum.

\begin{figure*}[!tp]\centering
\subfloat[Andersson--Onozawa (1996)]{\includegraphics[width=.491\textwidth,trim=0 6bp 0 7bp,clip]{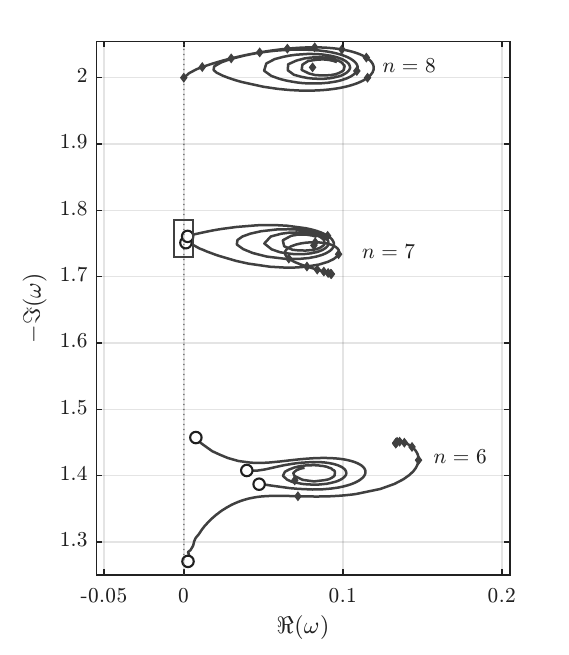}}\hfill
\subfloat[Present analytic continuation]{\includegraphics[width=.491\textwidth,trim=0 6bp 0 7bp,clip]{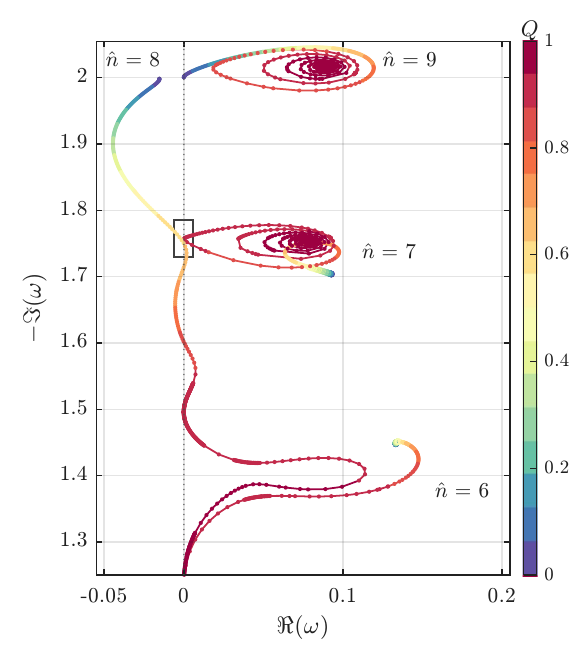}}\par\smallskip
\subfloat[Original paths near the imaginary axis]{\includegraphics[width=.491\textwidth,trim=0 6bp 0 7bp,clip]{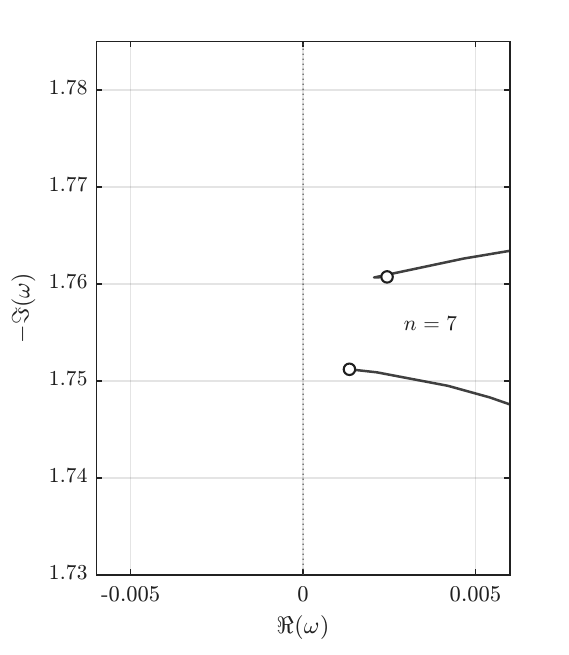}}\hfill
\subfloat[Continued paths near the imaginary axis]{\includegraphics[width=.491\textwidth,trim=0 6bp 0 7bp,clip]{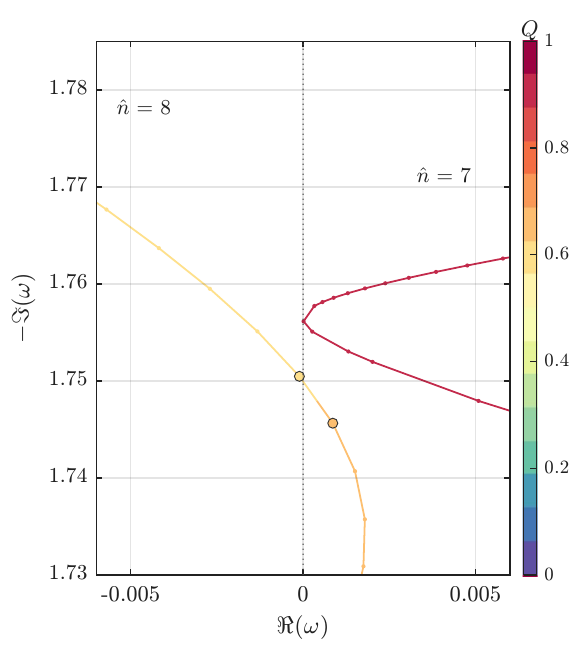}}
\caption{Gravitational $\ell=2$ trajectories from Andersson and Onozawa \cite[Fig.~2]{Andersson:1996xw} (left) and this work (right). At low charge their $n=6,7,8$ correspond to our $\hat n=6,7,9$, respectively. Our $\hat n=8$ denotes the unconventional Schwarzschild mode. The lower panels enlarge the same region. On the left, open circles mark endpoints and diamonds mark charge steps of $0.1$. On the right, symbols follow Fig.~\ref{fig:summary42l2}. The outlined points in (d) bracket the $\hat n=8$ crossing.}
\label{fig:RN_AO_comparison}
\end{figure*}

At $\ell=2$, Fig.~\ref{fig:RN_AO_comparison} resolves two paths left open near the imaginary axis in Ref.~\cite{Andersson:1996xw}. The $\hat n=7$ path turns. The $\hat n=8$ path crosses the axis between the two charge samples below.
\[
\begin{aligned}
\Re(\omega) &= -1.0832\times10^{-4} && (Q=0.622374),\\
\Re(\omega) &= +8.6060\times10^{-4} && (Q=0.627374).
\end{aligned}
\]

\subsection{QNM resonances and excitation factors}\label{sec:modeinteraction}\label{sec:excitation}

\begin{figure*}[!htbp]\centering
\subfloat[$\ell=2$, $\hat n=6,8$]{\includegraphics[width=.49\textwidth]{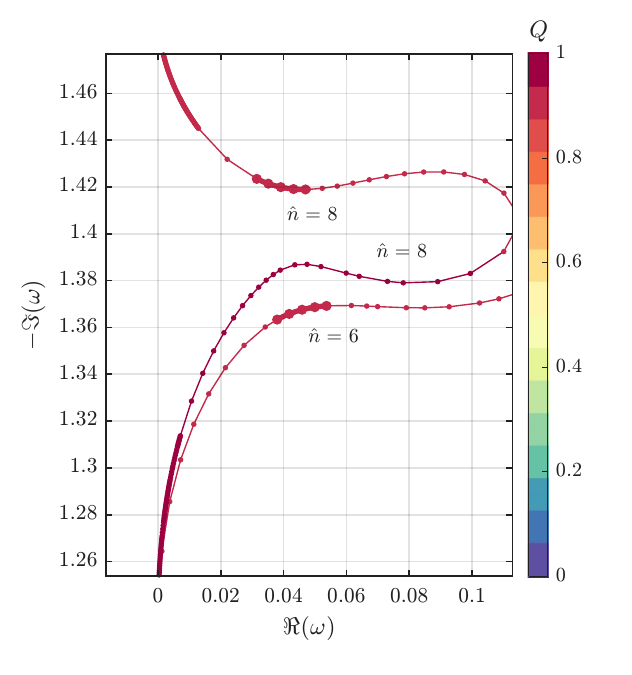}}\hfill
\subfloat[$\ell=3$, $\hat n=21$--28]{\includegraphics[width=.49\textwidth]{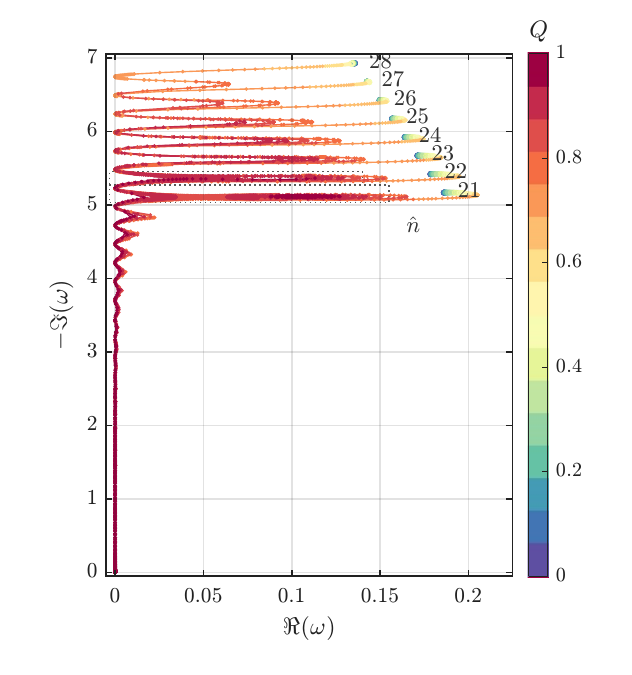}}
\caption{Overtone encounters at $\ell=2$ and $3$. Of the six pairs tested in (b), four show clear resonant enhancement and one has only shallow paired peaks. Color denotes $Q$.}
\label{fig:pairapproach}
\end{figure*}

\begin{figure*}[!tp]\centering
\includegraphics[width=.424\textwidth]{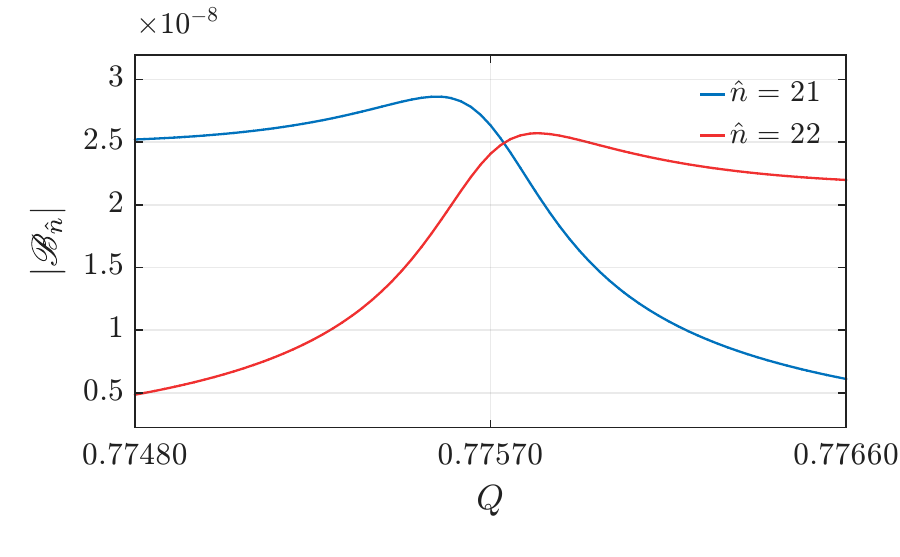}\hspace{.035\textwidth}
\includegraphics[width=.424\textwidth]{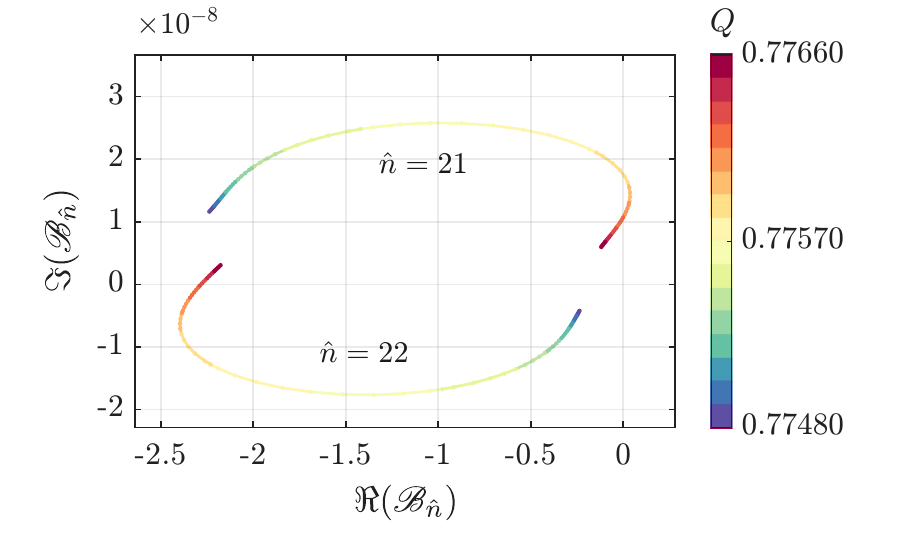}\par
\includegraphics[width=.424\textwidth]{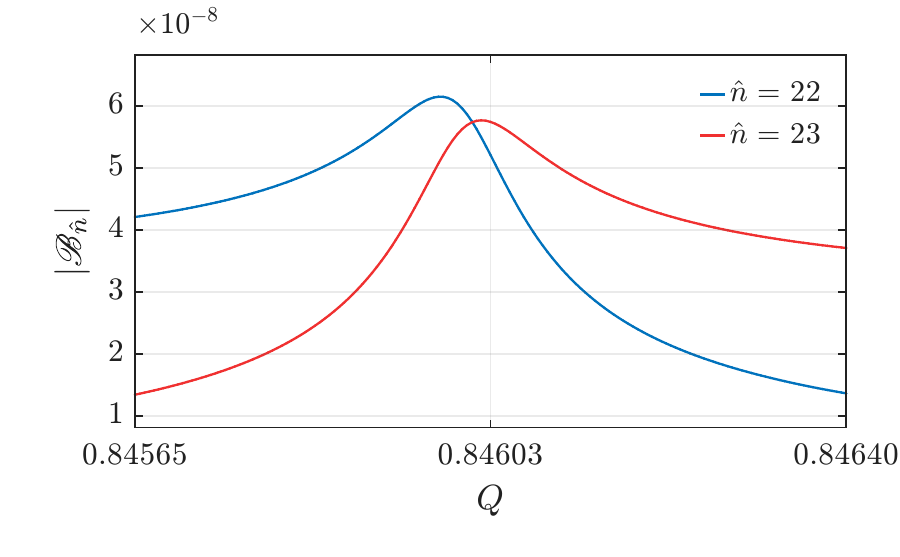}\hspace{.035\textwidth}
\includegraphics[width=.424\textwidth]{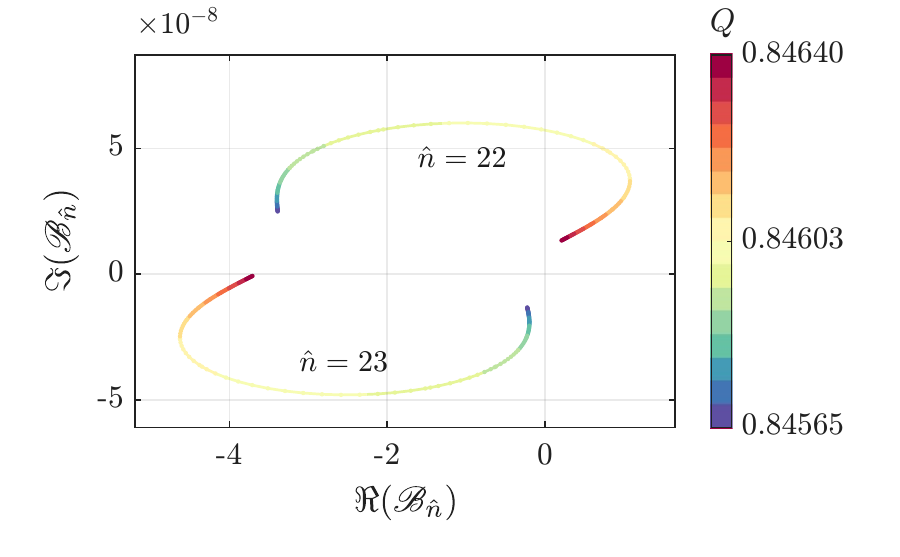}\par
\includegraphics[width=.424\textwidth]{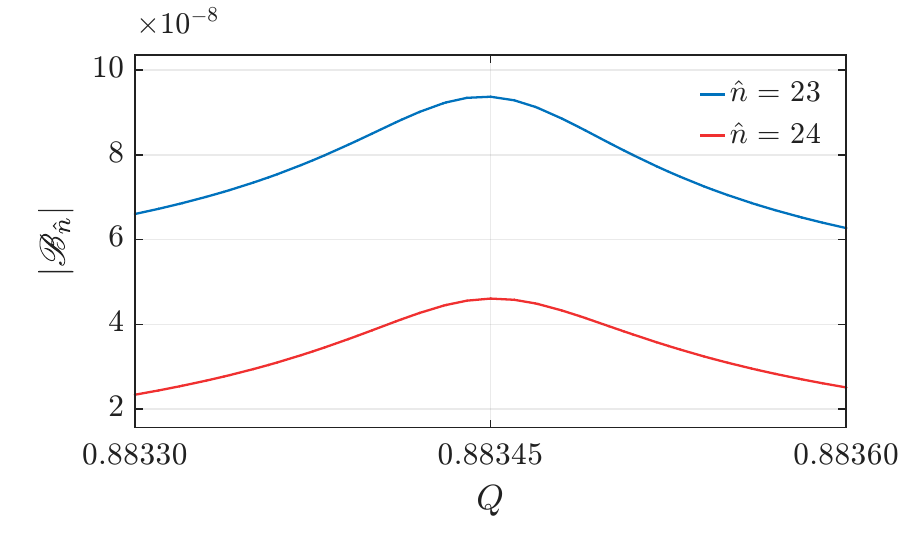}\hspace{.035\textwidth}
\includegraphics[width=.424\textwidth]{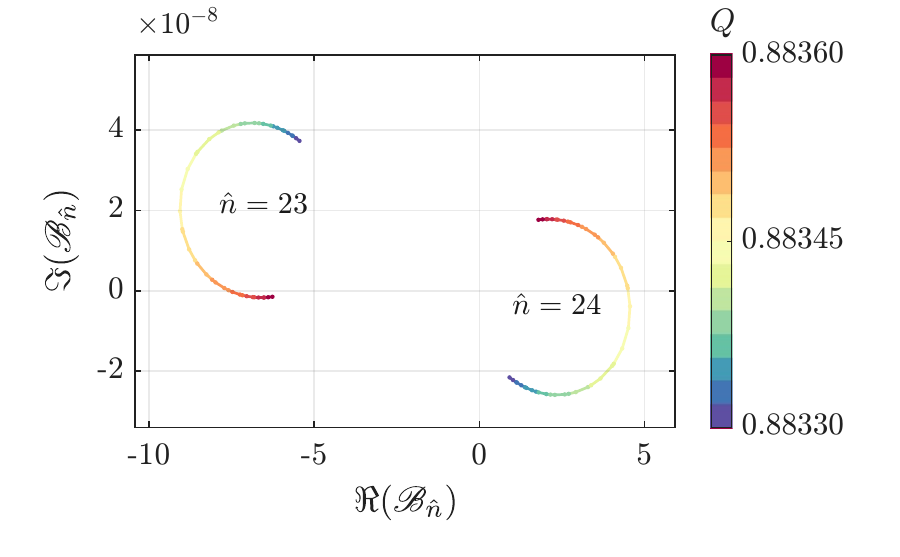}\par
\includegraphics[width=.424\textwidth]{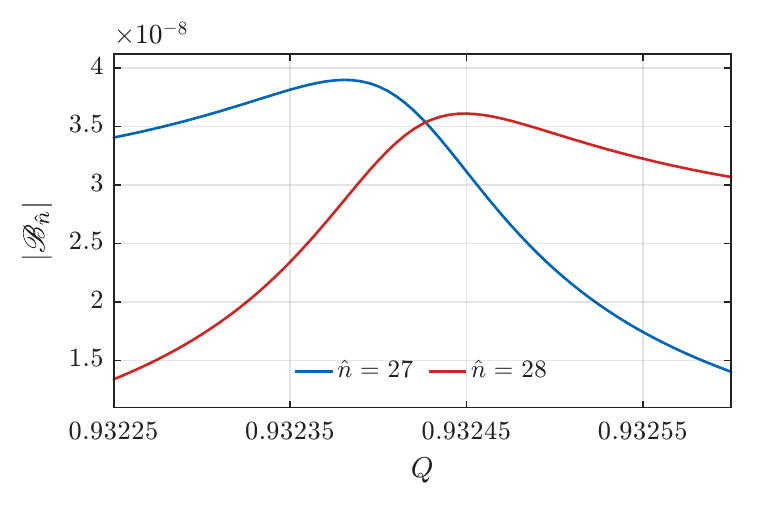}\hspace{.035\textwidth}
\includegraphics[width=.424\textwidth]{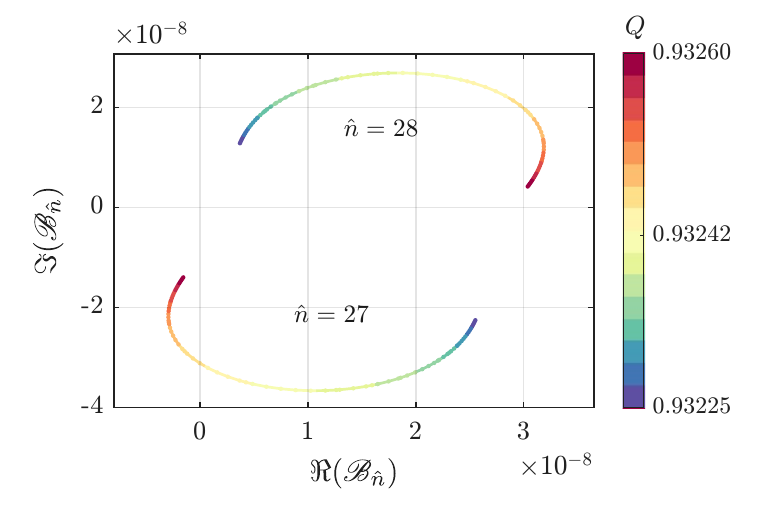}
\caption{Excitation factors of the resonant $\ell=3$ pairs $(21,22)$, $(22,23)$, $(23,24)$, and $(27,28)$, from top to bottom. The left panels give the magnitudes versus charge. The right panels trace the complex factors over the same intervals, colored by $Q$.}

\label{fig:excitationE}\label{fig:excitationP}\label{fig:excitationA}\label{fig:excitation}\label{fig:excitationlocal}\label{fig:excitationcomplexlocal}
\end{figure*}

Figure~\ref{fig:pairapproach}(b) identifies six pairs of nearby $\ell=3$ overtones. Figure~\ref{fig:excitation} shows the four pairs with clear resonant enhancement. Figure~\ref{fig:bandencounters} resolves the frequency trajectories in the two damping bands. The strongest peaks occur for $(22,23)$ near $Q=0.846$ and $(23,24)$ near $Q=0.88345$. For $(27,28)$ the two peaks bracket the minimum frequency spacing at $Q=0.932415$.

In the axial master normalization, the excitation factor at a simple root is \cite{Motohashi:2024fwt}

\begin{equation}
 \mathscr{B}_{\hat n}=\left.\frac{B^{\rm ref}}{2\omega\,\partial_\omega B^{\rm inc}}\right|_{\omega=\omega_{\hat n}}.
 \label{eq:excitationfactor}
\end{equation}

We evaluate the derivative along the real and imaginary frequency directions at fixed $Q$, radial homotopy and horizon normalization. The outgoing amplitude follows from the same horizon solution,

\begin{equation}
 B^{\rm ref}=\lim_{\rho\to+\infty}\frac{h(\rho){\rm{e}}^{-2\ii\omega r_*(\rho)}}{g_{\infty}^{\mathrm{up}}(1/\rho)}.
 \label{eq:outgoingamplitude}
\end{equation}

The factor is unchanged by an analytic nonzero rescaling of the horizon solution. For $(\partial_{r_*}^2+\omega^2-V)G_\omega=\delta(r_*-r_*')$, it also determines the Green function pole residue:

\begin{equation}
 \mathop{\rm Res}_{\omega=\omega_{\hat n}}G_\omega(r,r')
 =\frac{\mathscr{B}_{\hat n}}{\ii}\,R^{\infty}_{\mathrm{up}}(r,\omega_{\hat n})R^{\infty}_{\mathrm{up}}(r',\omega_{\hat n}).
 \label{eq:excitationresidue}
\end{equation}

For the two strongest pairs, $|\Delta\mathscr{B}|$ follows the inverse frequency spacing $1/|\Delta\omega|$ near closest approach. Both differences are taken between the two overtones at the same charge.

These excitation factors characterize the poles of the Green function. Their contributions to a waveform also depend on the source integrals and on interference between modes.

\subsection{Near-extremal and extremal spectra}\label{sec:extremalresults}

Of the 84 families, 75 approach distinct finite-damping extremal roots. Each reaches at least $Q=0.99999$. The largest separation between a final subextremal frequency and its extremal root is approximately $1.15\times10^{-6}$. At $Q=0.995$ and $0.999$, the largest incoming-amplitude residual over these families and both extraction radii is $1.95\times10^{-19}$ for $h(0)=1$. Under changes in the horizon start, expansion order, ray angle and extraction radii, the largest measured frequency shift is approximately $9.51\times10^{-21}$. The remaining nine families approach zero frequency. Table~\ref{tab:zdm} gives their damping rates and excitation magnitudes down to $1-Q=10^{-14}$.

For the accumulating sequences, introduce
\begin{subequations}\label{eq:neparameters}
\begin{align}
 \epsilon&=\sqrt{1-Q^2},\qquad
 \sigma=\frac{r_+-r_-}{r_+}=\frac{2\epsilon}{1+\epsilon},\\
 \kappa_+&=\frac{\epsilon}{(1+\epsilon)^2}.
\end{align}
\end{subequations}

The gravitational near-extremal prediction is \cite{Zimmerman:2015trm}
\begin{equation}
 \omega_{\ell p}^{\rm ZDM}
   =-\ii\kappa_+(p+\ell)+O(\sigma^2),
 \qquad p=0,1,\ldots,
 \label{eq:gemzdm}
\end{equation}

The amplitude zeros give $-\Im(\omega)/\kappa_+\to p+\ell$, with $p=0,1$ for $\ell=2$ and $p=0,\ldots,6$ for $\ell=3$. In the sampled near-extremal interval the departure from this leading frequency is consistent with a correction of order $\sigma^2$. At $1-Q=10^{-14}$, varying the working precision and radial integration settings reproduced all nine damping rates to the 12 significant digits quoted in Table~\ref{tab:zdm}.

\onecolumngrid
\noindent\begin{minipage}[b]{\textwidth}
\makeatletter\def\@captype{table}\makeatother
\centering\small
\caption{Damping and excitation of the nine families approaching zero frequency.}
\label{tab:zdm}\label{tab:zdmdeep}\label{tab:originexcitation}
\begin{tabular*}{.99\textwidth}{@{\extracolsep{\fill}}ccrrrr@{}}\toprule
& & \multicolumn{2}{c}{$-\Im(\omega)$} & \multicolumn{2}{c}{$|\mathscr{B}_{\hat n}|$} \\
\cmidrule(lr){3-4}\cmidrule(lr){5-6}
$\ell$ & $\hat n$ & $Q=0.9999$ & $Q=1-10^{-14}$ & $Q=0.9999$ & $Q=1-10^{-14}$\\\midrule
2 & 6 & $0.02829827$ & $2.82842712475\times10^{-7}$ & $1.247\times10^{-15}$ & $1.931\times10^{-60}$ \\
2 & 8 & $0.04246267$ & $4.24264068712\times10^{-7}$ & $3.206\times10^{-14}$ & $5.865\times10^{-59}$ \\
3 & 21 & $0.04244450$ & $4.24264068712\times10^{-7}$ & $3.762\times10^{-23}$ & $7.507\times10^{-88}$ \\
3 & 22 & $0.05660495$ & $5.65685424949\times10^{-7}$ & $1.411\times10^{-21}$ & $3.374\times10^{-86}$ \\
3 & 24 & $0.07077152$ & $7.07106781187\times10^{-7}$ & $1.977\times10^{-20}$ & $5.631\times10^{-85}$ \\
3 & 25 & $0.08494419$ & $8.48528137424\times10^{-7}$ & $1.593\times10^{-19}$ & $5.381\times10^{-84}$ \\
3 & 26 & $0.09912296$ & $9.89949493661\times10^{-7}$ & $8.911\times10^{-19}$ & $3.562\times10^{-83}$ \\
3 & 27 & $0.11330778$ & $1.13137084990\times10^{-6}$ & $3.838\times10^{-18}$ & $1.814\times10^{-82}$ \\
3 & 28 & $0.12749863$ & $1.27279220614\times10^{-6}$ & $1.357\times10^{-17}$ & $7.585\times10^{-82}$ \\
\bottomrule\end{tabular*}\par
\end{minipage}\par
\vspace{\baselineskip}
\twocolumngrid

As the nine frequencies approach zero, their individual excitation factors also vanish. Over $10^{-14}\le 1-Q\lesssim10^{-12}$, the measured magnitudes scale as $(\kappa_+)^9$ for $\ell=2$ and $(\kappa_+)^{13}$ for $\ell=3$. At $1-Q=10^{-14}$ the relative change in each excitation factor under changed radial settings is below $9.38\times10^{-13}$.

The finite-damping endpoints of the families traced here all lie on the conventional sheet. For the first forty-two families at each angular order studied, we find no endpoint at an isolated unconventional-sheet QNM of the kind found for extremal Kerr \cite{Chen:2026extremal}. Extremal RN itself does admit gravitational QNMs on the unconventional sheet. Table~\ref{tab:negativefinite} gives an example at each angular order. At these frequencies the incoming amplitude vanishes on continuation across the negative imaginary axis and is nonzero on the conventional sheet. Their Schwarzschild parents are unassigned. Table~\ref{tab:xrall} lists the extremal limit of each overtone. Tables~\ref{tab:latestfinite2} and~\ref{tab:latestfinite3} give the last computed subextremal frequencies for the finite-damping families.

\section{Conclusions}\label{sec:conclusions}

We have computed the RN gravitational spectrum from zero charge to extremality for $\ell=2,3$ and $\hat n=0,\ldots,41$. Analytic continuation lets us follow a frequency through the negative imaginary axis and identify its Schwarzschild parent. Eight consecutive $\ell=3$ overtones cross the axis. Near the $\ell=2$ Schwarzschild AS frequency, the unconventional mode $\hat n=8$ crosses the axis and approaches zero frequency, whereas the neighboring $\hat n=9$ branch retains finite damping. The scattering amplitudes distinguish the exact special solutions at the two endpoints. The polar Schwarzschild AS mode satisfies the QNM condition, while the axial AS modes at $Q=1$ have zero reflection and nonzero incident amplitude and hence describe total-transmission modes.

Every one of the 84 families has an extremal limit. Seventy-five approach distinct roots computed directly at $Q=1$ on the conventional sheet. The other nine families accumulate at zero frequency. In the sampled near-extremal range their excitation magnitudes fall as $(\kappa_+)^9$ for $\ell=2$ and $(\kappa_+)^{13}$ for $\ell=3$. A separate search at $Q=1$ also gives a gravitational QNM on the unconventional sheet for each angular order. At intermediate charge, the two strongest resonances involve the $\ell=3$ pairs $(22,23)$ and $(23,24)$, for which the magnitude of the difference between the excitation factors grows as the frequency spacing narrows.

This accounts for all 42 gravitational families at each of the two angular orders and resolves both questions posed by Berti~\cite{Berti:2004md,Berti:2009kk} for the RN spectrum in this range. The same frequencies provide initial data for a calculation of the Kerr--Newman spectrum. At fixed charge, each RN frequency can seed a continuation in spin. Its overtone label then identifies the nonrotating parent of the resulting branch, giving a systematic route to the complete spectrum of a rotating, charged black hole.

\section*{Data availability}\label{sec:data-availability}
The numerical data for this study will be released in the \href{https://github.com/IronChen1/Complete-QNMs}{Complete-QNMs} repository~\cite{ChenQNM}.

\section*{Acknowledgments}
This work was supported by the National Key Research and Development Program of China (No.~2021YFC2203001) and the National Natural Science Foundation of China (No.~12475049).

\appendix
\section{Series construction and radial analytic transport}\label{app:series}

\subsection{Subextremal horizon recurrence}

For $Q<1$, insert $R_\ell^{\mathrm{in}}={\rm{e}}^{-\ii\omega r_*}h$ into Eq.~\eqref{eq:master}. Differentiating the phase cancels the $\omega^2h$ term, leaving $\partial_{r_*}^2h-2\ii\omega\partial_{r_*}h-Vh=0$. With $x=(r-r_+)/(r-r_-)$, use $\partial_{r_*}=f(\partial_r x)\partial_x$ and clear the denominators. The resulting equation, with primes denoting $x$ derivatives, is

\begin{equation}
 D(x)h''+E(x)h'+F(x)h=0,
 \label{eq:hode}
\end{equation}

\begin{subequations}\label{eq:polys}
\begin{align}
 D&=x(r_+-r_-x)^2(1-x)^2,\\
 E&=-2x(r_+-r_-x)^2(1-x)\nonumber\\
  &\quad+(r_++r_-x)(r_+-r_-x)(1-x)^2\nonumber\\
  &\quad-\frac{2\ii\omega(r_+-r_-x)^4}{r_+-r_-},\\
 F&=-L(r_+-r_-x)^2+q_+(r_+-r_-x)(1-x)\nonumber\\
  &\quad-4Q^2(1-x)^2 .
\end{align}
\end{subequations}

At the outer horizon $x=0$, the ingoing normalization fixes $a_0=1$. Inserting $h=\sum_{j\ge0}a_jx^j$ into Eq.~\eqref{eq:hode} and collecting $x^{m-1}$ gives the recurrence below, with negative-index terms set to zero.

\begin{equation}
 a_m=-\frac{\mathcal R_m}{m[D_1(m-1)+E_0]},\qquad m\geq1 ,
 \label{eq:rec}
\end{equation}

\begin{align}
 \mathcal R_m={}&
 \sum_{k=2}^{\deg D}D_k(m+1-k)(m-k)a_{m+1-k}\nonumber\\
 &+\sum_{k=1}^{\deg E}E_k(m-k)a_{m-k}
 +\sum_{k=0}^{\deg F}F_k a_{m-1-k}.
 \label{eq:rhs}
\end{align}
Here $D_k,E_k,F_k$ denote the polynomial coefficients. The nearest singularity to the horizon in the $x$ plane is at $x=1$; the coefficients therefore furnish a convergent local initialization for $|x_0|<1$, away from exceptional frequency normalizations. The derivative is evaluated from the same series, not by numerical differencing.

With $\eta_{\rm H}=\ii\omega/\kappa_+$, the denominator is $mr_+^2(m-\eta_{\rm H})$. The recurrence degenerates when $\eta_{\rm H}=m\ge1$. If $\mathcal R_m=0$ at that order, continuation in $\omega$ fixes the undetermined coefficient $a_m$.

\subsection{Infinity asymptotic coefficients}

At infinity the series in Eq.~\eqref{eq:directbranches} are asymptotic. We truncate where the largest of four consecutive term magnitudes is smallest. The four-term window prevents an isolated small coefficient from setting the truncation order. We check the result by varying the expansion settings.

With $z=1/r$, write $g_{\infty}^{\mathrm{up}}(z)\sim\sum b_nz^n$ with $b_0=1$. The coefficients obey

\begin{align}
 2\ii\omega(n+1)b_{n+1}
 ={}&[n(n+1)-L]b_n\nonumber\\
 &+[q_+-2(n-1)(n+1)]b_{n-1}\nonumber\\
 &+Q^2[(n-2)(n+1)-4]b_{n-2}.
 \label{eq:outrec}
\end{align}
The incoming recurrence follows by $\omega\mapsto-\omega$. In the primary QNM calculation, this recurrence supplies the infinity denominator directly, so only the horizon solution is propagated.

\subsection{Convergent transport and large charge conditioning}\label{app:transport}

For propagation to large radius we return to $r$, since $x$ approaches unity at infinity. Applying $\partial_{r_*}=f\partial_r$ to the phase-factored equation gives $f\partial_r^2h+(\partial_r f-2\ii\omega)\partial_rh-(V/f)h=0$. Multiplication by $r^4$ and substitution of Eq.~\eqref{eq:potential} give

\begin{align}
 &(r^4-2r^3+Q^2r^2)\partial_r^2h\nonumber\\
 &\quad+(2r^2-2Q^2r-2\ii\omega r^4)\partial_rh\nonumber\\
 &\quad+(-Lr^2+q_+r-4Q^2)h=0 .
 \label{eq:directrode}
\end{align}
We propagate $(h,\partial_rh)$ in overlapping Taylor disks, limiting the step by the distance to singularities, the local phase scale and the series tail. Both components share the same rescaling, which we undo to recover the scattering amplitude. From the real exterior the standard contour enters the upper half of the $r$ plane through the image of $x_b=0.6+0.35\ii$ under $r(x)=(r_+-r_-x)/(1-x)$. This intermediate point lies inside $|x|<1$. The next point, $3{\rm{e}}^{\ii\theta}$, joins the extraction ray at a radius larger than both horizon radii ($r_+\le2$ in our units). The contour then follows the ray outward.

For a trial frequency $\omega$, Eq.~\eqref{eq:directangle} requires $\sin(\arg_{\rm c}\omega+\theta)>0$ so that the outgoing wave decays relative to the incoming wave. For strongly damped modes $\arg_{\rm c}\omega$ is near $-\pi/2$, placing the direction of fastest asymptotic decay near $\pi$. We use $\pi-0.15$ radians to stay in the upper half-plane. For weakly damped modes with positive real frequency, $2$ radians lies inside the decay sector. We keep the angle fixed during each local solve and check the suppression in Eq.~\eqref{eq:directsuppression} at finite radius.

To select the real starting point $r_0$ near coalescing horizons, we evaluate the tortoise coordinate there in terms of $x_0$:

\begin{subequations}\label{eq:starthorizonconditioning}
\begin{align}
 r_*(r_0)&=r_0+\frac{r_+^2}{r_+-r_-}\ln x_0
       +2\ln\frac{r_+-r_-}{1-x_0},\\
 r_0&=\frac{r_+-r_-x_0}{1-x_0}.
\end{align}
\end{subequations}
We choose the horizon start adaptively within $0<x_0<1$, screening the exponential factor with $|{\rm{e}}^{2\ii\omega r_*(r_0)}|\geq10^{-12}$ in the stated tortoise normalization. Horizon tails and summation cancellation are checked separately, and the series length is increased where necessary at fixed working precision. The cutoff limits suppression of the competing horizon solution at the start to twelve orders of magnitude. We test this numerical choice by varying the start and propagation settings.

As the horizons merge, $r(x_b)-r_+=(r_+-r_-)x_b/(1-x_b)$ shrinks to zero. A contour through $r(x_b)$ would bring an outward-shifted start back toward the horizon. We therefore choose an intermediate point directly in $r$, with a fixed positive imaginary part, and use the path

\begin{equation}
 r_0\ \longrightarrow\ 1.7+0.7\ii\ \longrightarrow\ 3{\rm{e}}^{\ii\theta}
 \ \longrightarrow\ \rho {\rm{e}}^{\ii\theta} .
 \label{eq:outerbridge}
\end{equation}
Here $1.7+0.7\ii$ stays a finite distance from the merging horizons and joins the same upper contour without extra winding. In a near-extremal stability check we used $1.7+1.2\ii$, reduced $\theta$ by $0.025$ radians and varied the extraction radii. We evaluate the incoming-amplitude residual at two radii on the ray.

\subsection{The irregular horizon at exact extremal charge}\label{app:extremalconstruction}

At $Q=1$ the horizons coincide at $r=1$ and the subextremal coordinate degenerates. We use $x=(r-1)/r=1-1/r$, which places the horizon at $x=0$ and infinity at $x=1$. In Eq.~\eqref{eq:potential} the square root in $q_+$ reduces to $2\ell+1$, giving $q_+=2\ell+4$. Setting $Q=1$ in Eq.~\eqref{eq:directrode} and using $\partial_r=(1-x)^2\partial_x$ gives

\begin{align}
 &x^2(1-x)^2 h''
   +(-2\ii\omega+2x-6x^2+4x^3)h'\nonumber\\
 &\quad+\big[-\ell(\ell-1)+(4-2\ell)x-4x^2\big]h=0 .
 \label{eq:xrhode}
\end{align}

Substituting the formal horizon series $h\sim\sum_{n\ge0}a_nx^n$ with $a_0=1$ and equating powers of $x$ gives

\begin{align}
2\ii\omega(n+1)a_{n+1}
={}&[n(n+1)-\ell(\ell-1)]a_n\nonumber\\
 &+[4-2\ell-2(n-1)(n+1)]a_{n-1}\nonumber\\
 &+[(n-2)(n+1)-4]a_{n-2}.
 \label{eq:xrrec}
\end{align}
Under inversion this is the outgoing infinity recurrence of the electromagnetic branch with angular order $\ell-1$. Except for terminating solutions the series is asymptotic. We truncate at the minimum of the term envelope. The leading horizon exponential $ {\rm{e}}^{\ii\omega/x}$ decays when $\Im(\omega/x)>0$. We start at the centre of this sector for the seed frequency by choosing

\begin{subequations}\label{eq:xrstart}
\begin{align}
 x_0&=\frac{{\rm{e}}^{-\ii\theta_{\rm H}}}{\rho_{\rm H}},\qquad
 \theta_{\rm H}=\frac{\pi}{2}-\arg_{\rm c}\omega_{\rm seed},\\
 \rho_{\rm H}&=\max\!\left(24,\frac{140}{|\omega|}\right).
\end{align}
\end{subequations}
The constants $24$ and $140$ keep $|x_0|\le1/24$ and $|\omega/x_0|\ge140$, respectively. These numerical settings control the starting distance and exponential scale. We first propagate to $r=1+0.35{\rm{e}}^{-\ii\theta_{\rm H}}$. The circle of radius $0.35$ about $r=1$ encloses only the horizon singularity. We follow its arc with continuous argument from $-\theta_{\rm H}$ to zero, then join the upper exterior contour in Eq.~\eqref{eq:outerbridge}. Keeping this continuation as $\Re(\omega)$ crosses zero gives the unconventional-sheet amplitude. By reality of the potential, the conventional-sheet amplitude there is the complex conjugate of the upper-contour amplitude at $-\overline{\omega}$.

Propagation then uses convergent ordinary-point expansions. We find the zeros of the incoming amplitude in Eq.~\eqref{eq:binclimit} using the extremal tortoise coordinate and infinity recurrence.

\onecolumngrid
\section{Frequency tables and trajectories}\label{app:frequencies}\label{app:atlas}

\begin{figure}[p]
\centering
\includegraphics[width=\textwidth,height=.913\textheight,keepaspectratio]{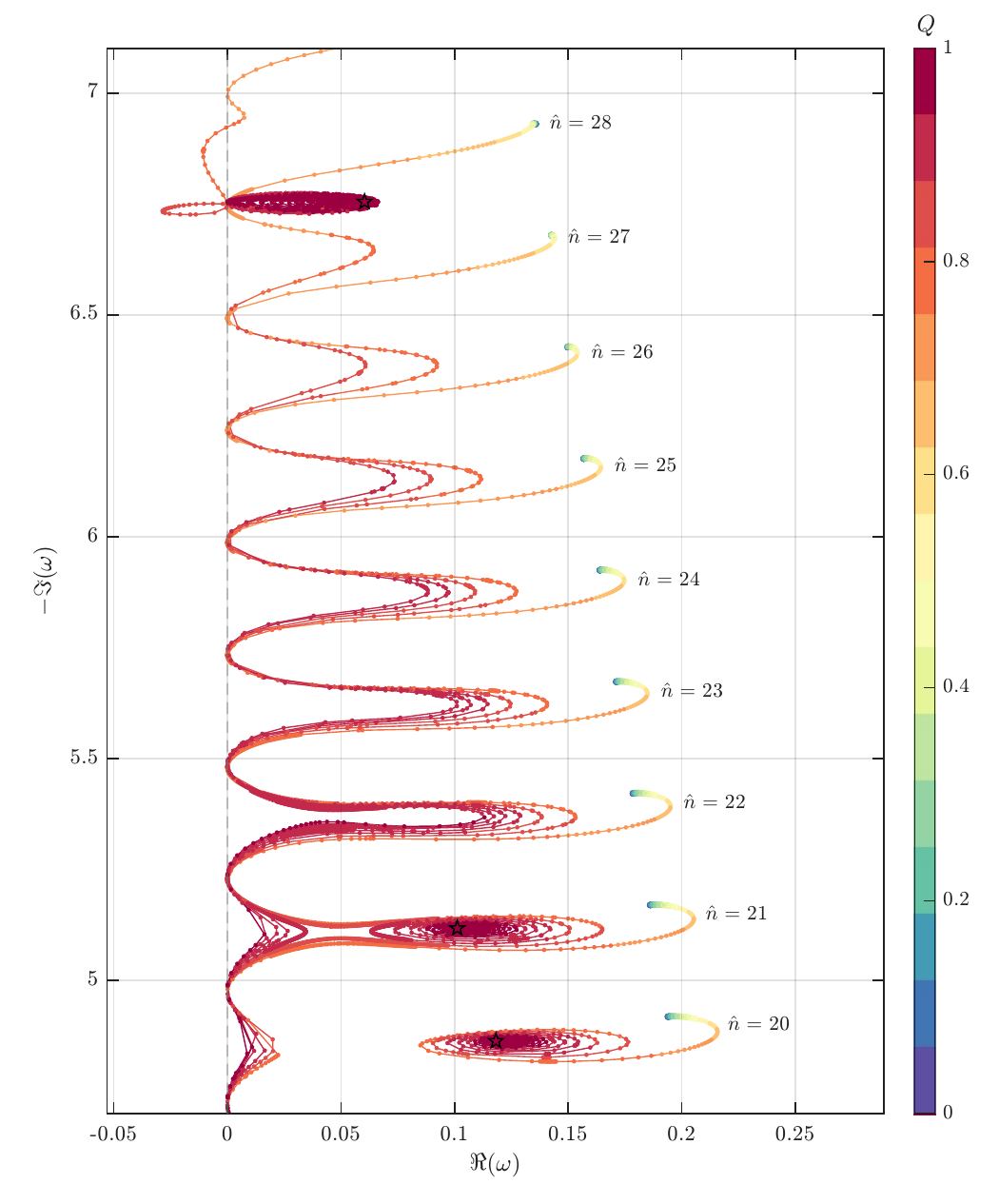}
\caption{Intermediate-damping spirals at $\ell=3$. The dashed line marks $\Re(\omega)=0$, labels mark the low-charge ends, and stars mark the $Q=1$ frequencies. Colors follow Fig.~\ref{fig:summary42l2}.}
\label{fig:bandstrict}
\end{figure}

\begin{figure}[p]
\centering
\setlength{\abovecaptionskip}{3pt}
\makebox[\textwidth][c]{\includegraphics[width=.49\textwidth]{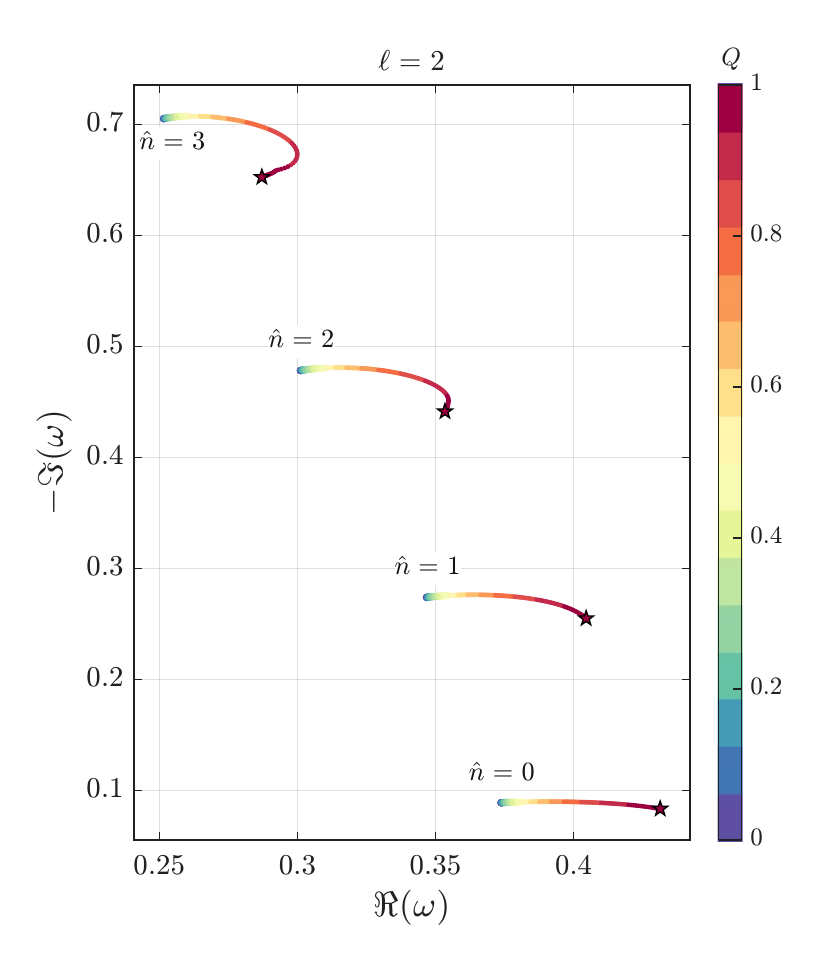}\hspace{.015\textwidth}\includegraphics[width=.49\textwidth]{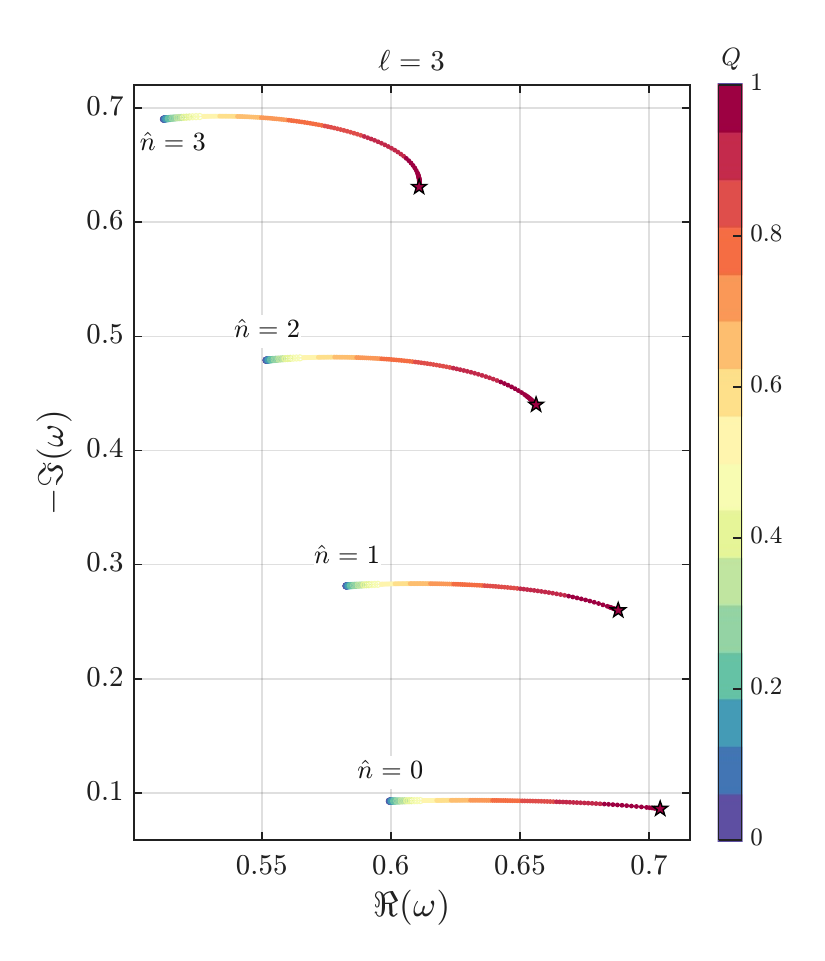}}\par\vspace{2pt}
\makebox[\textwidth][c]{\includegraphics[width=.49\textwidth]{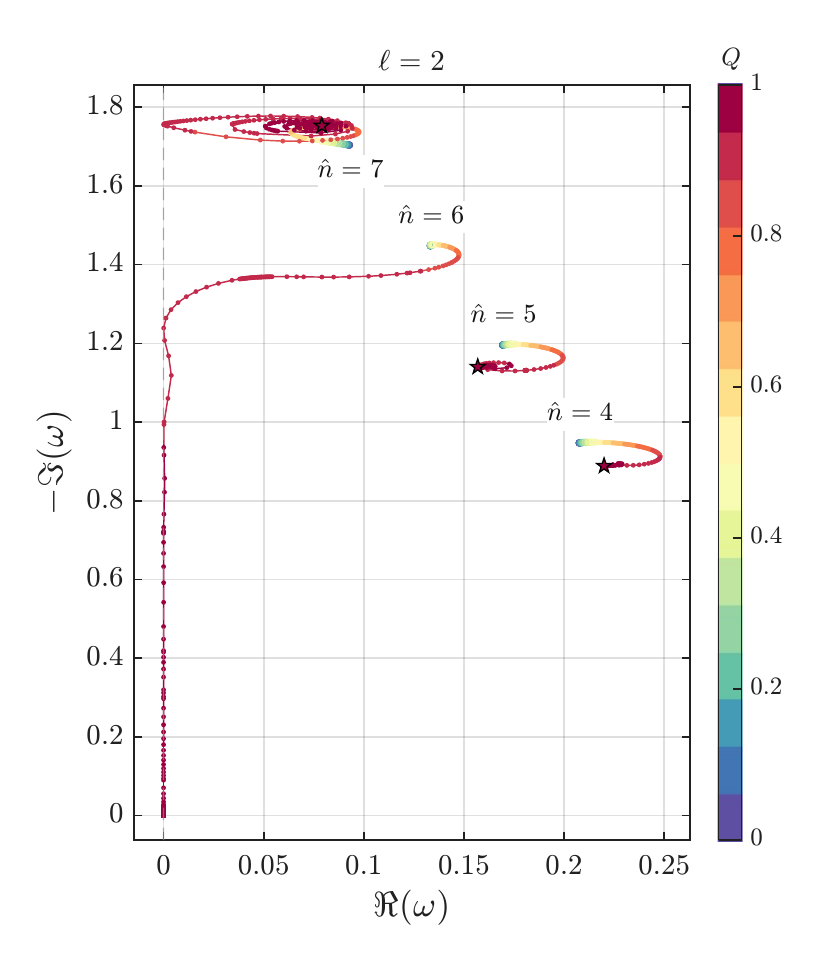}\hspace{.015\textwidth}\includegraphics[width=.49\textwidth]{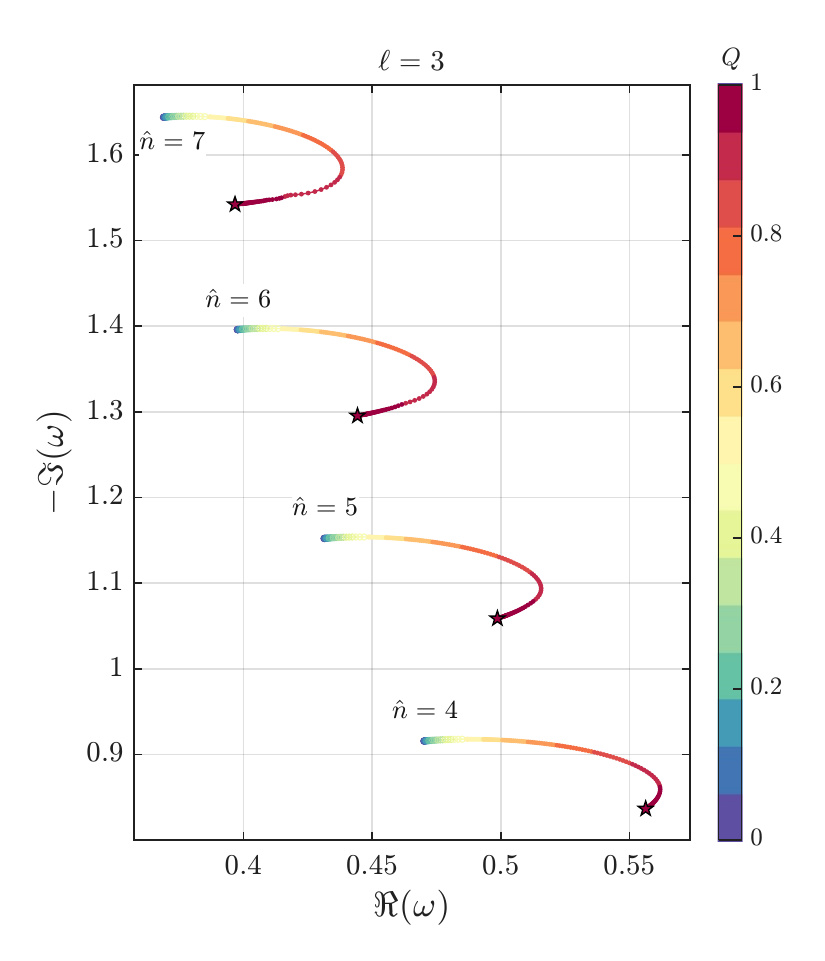}}
\caption{Overtones $\hat n=0$--$3$ (top) and $4$--$7$ (bottom) for $\ell=2$ (left) and $\ell=3$ (right). Symbols and colors follow Fig.~\ref{fig:summary42l2}.}
\label{fig:atlas1}
\end{figure}

\begin{figure}[p]
\centering
\setlength{\abovecaptionskip}{3pt}
\makebox[\textwidth][c]{\includegraphics[width=.49\textwidth]{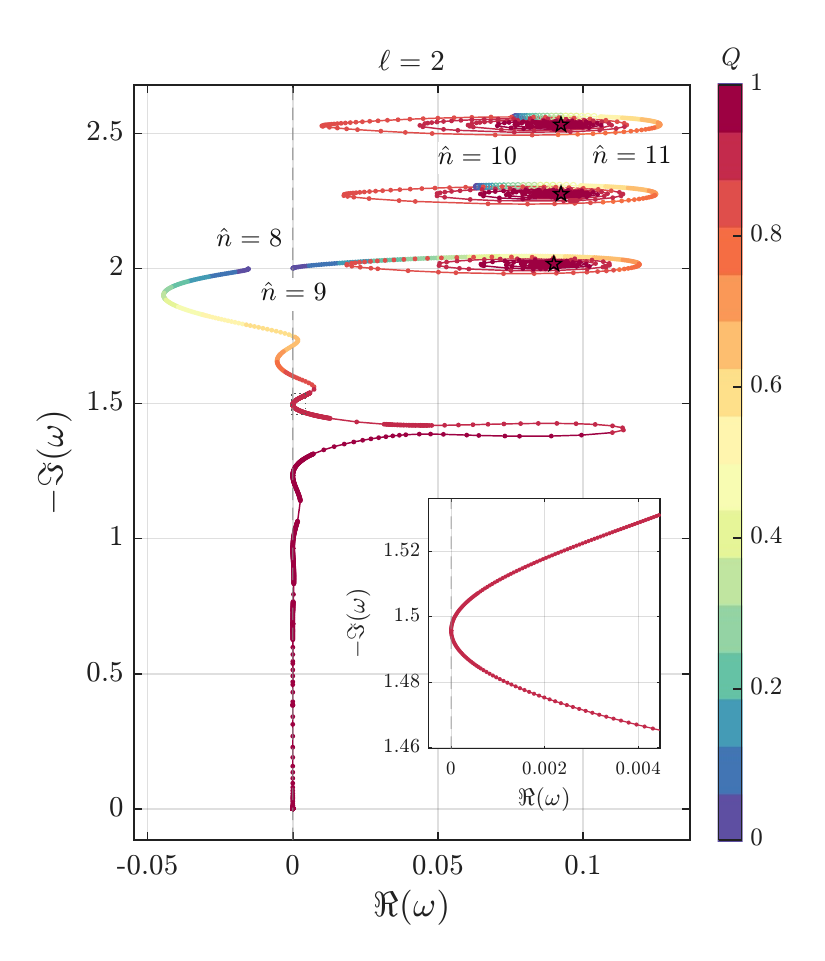}\hspace{.015\textwidth}\includegraphics[width=.49\textwidth]{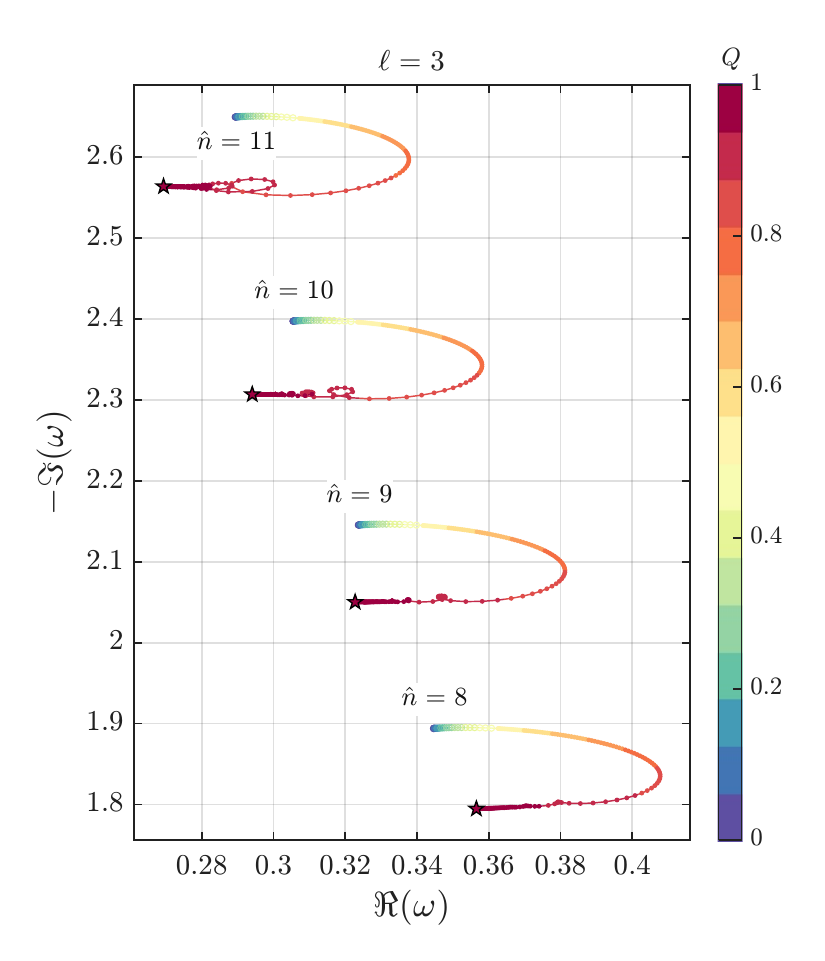}}\par\vspace{2pt}
\makebox[\textwidth][c]{\includegraphics[width=.49\textwidth]{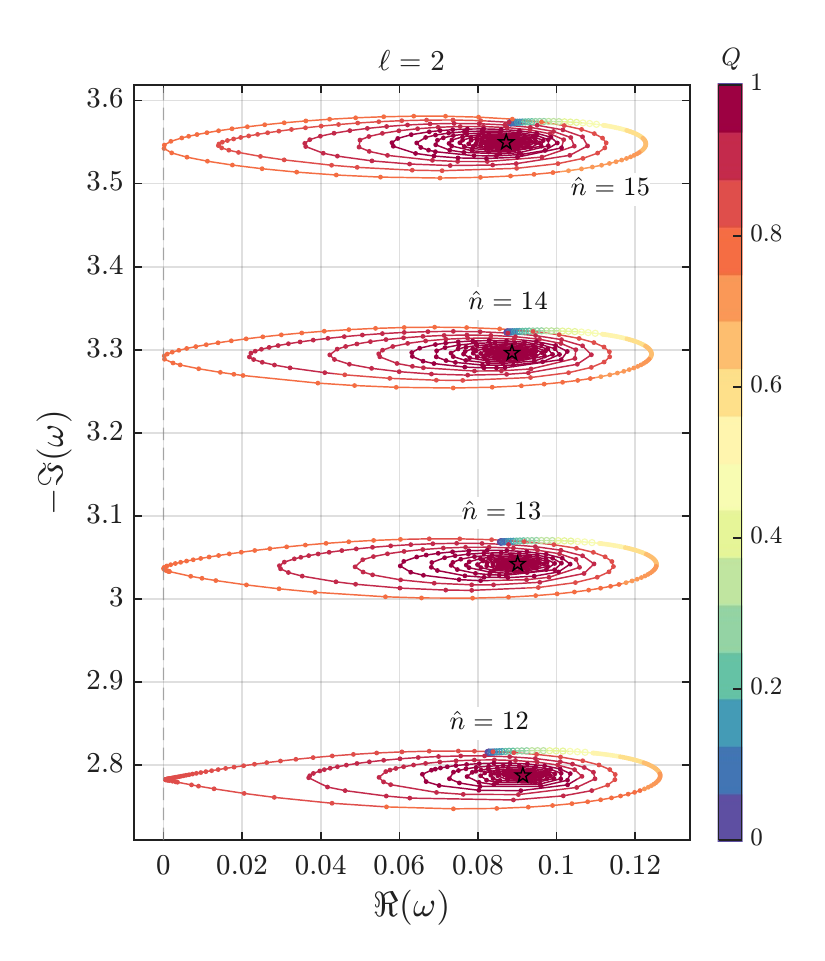}\hspace{.015\textwidth}\includegraphics[width=.49\textwidth]{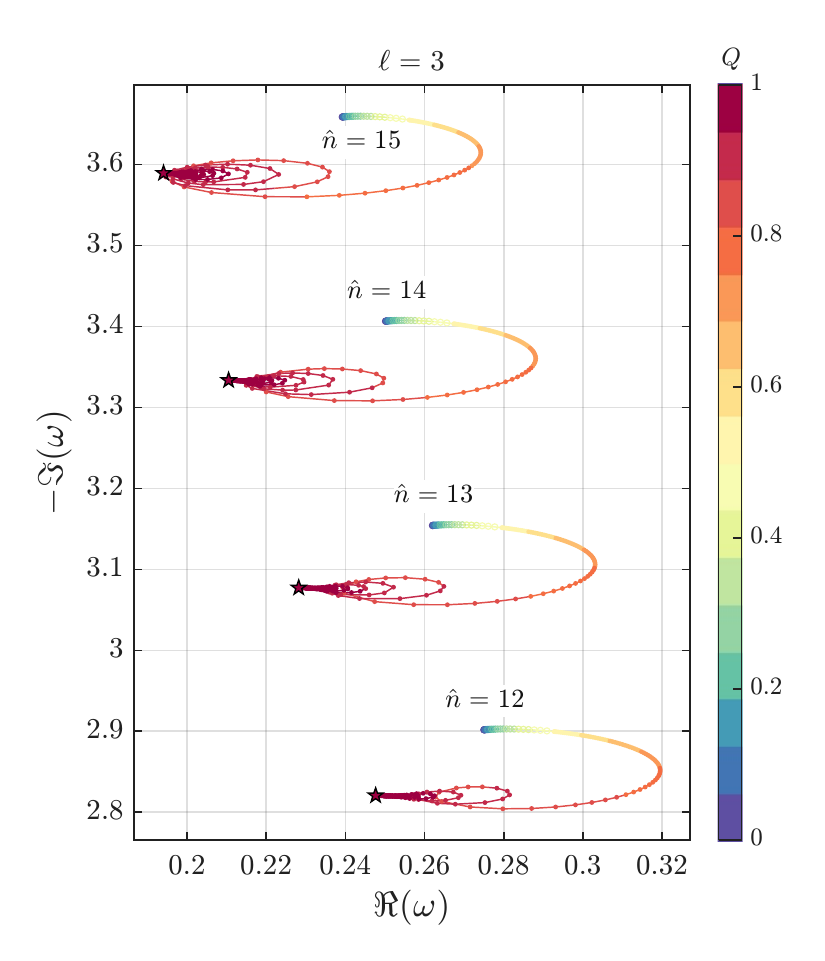}}
\caption{Overtones $\hat n=8$--$11$ (top) and $12$--$15$ (bottom). The inset follows $\ell=2$, $\hat n=8$ near the imaginary axis at $0.88\leq Q\leq0.90$.}
\label{fig:atlas2}
\end{figure}

\begin{figure}[p]
\centering
\setlength{\abovecaptionskip}{3pt}
\makebox[\textwidth][c]{\includegraphics[width=.49\textwidth]{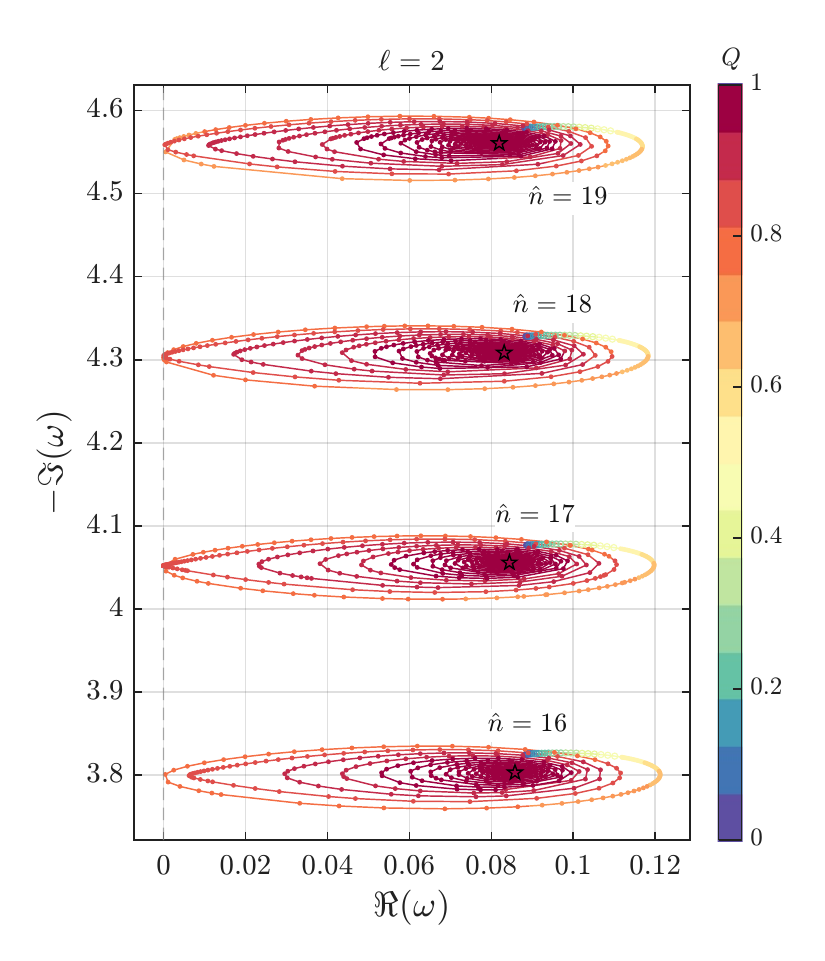}\hspace{.015\textwidth}\includegraphics[width=.49\textwidth]{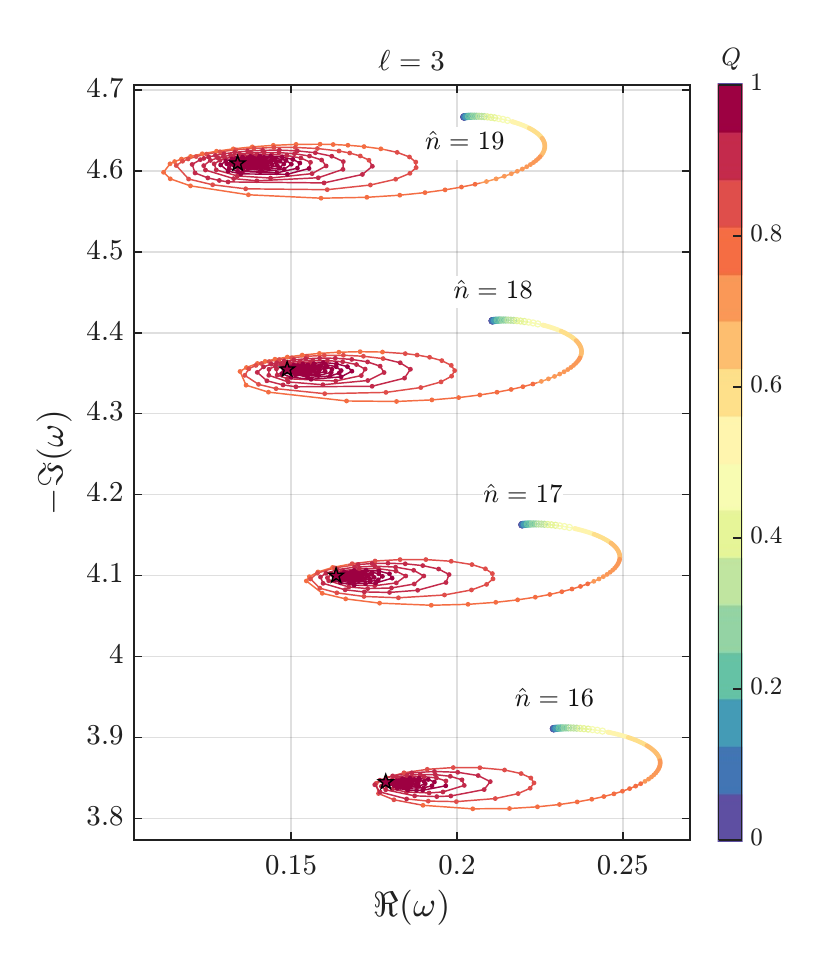}}\par\vspace{2pt}
\makebox[\textwidth][c]{\includegraphics[width=.49\textwidth]{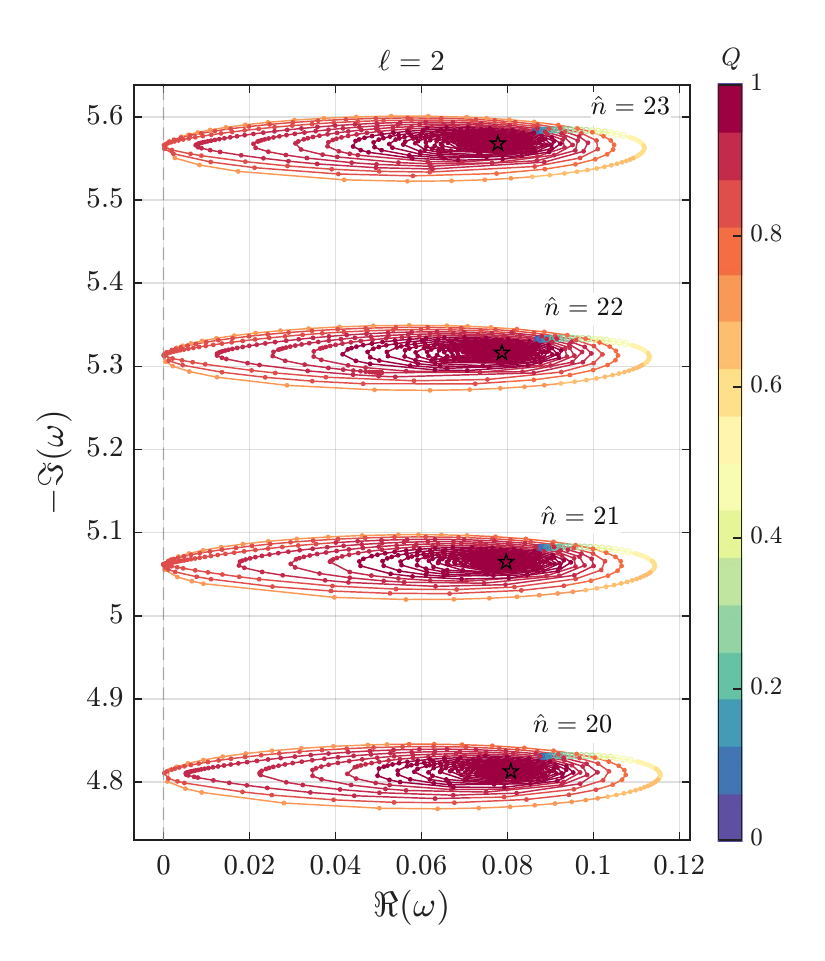}\hspace{.015\textwidth}\includegraphics[width=.49\textwidth]{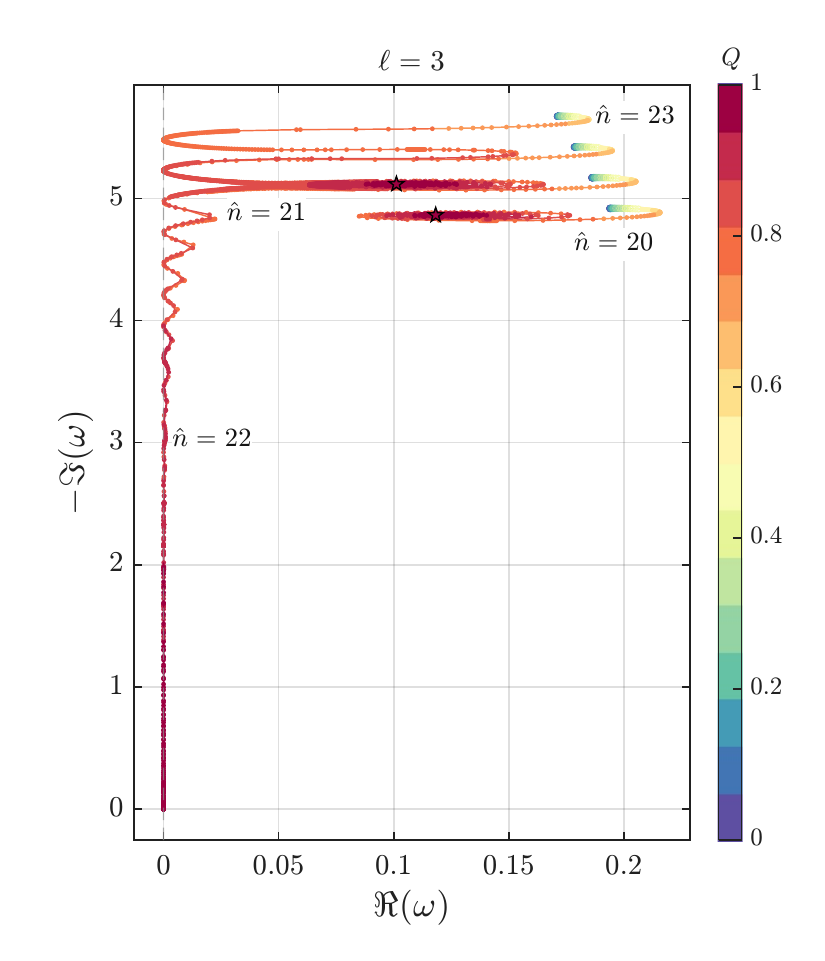}}
\caption{Overtones $\hat n=16$--$19$ (top) and $\hat n=20$--$23$ (bottom).}
\label{fig:atlas3}
\end{figure}

\begin{figure}[p]
\centering
\setlength{\abovecaptionskip}{3pt}
\makebox[\textwidth][c]{\includegraphics[width=.49\textwidth]{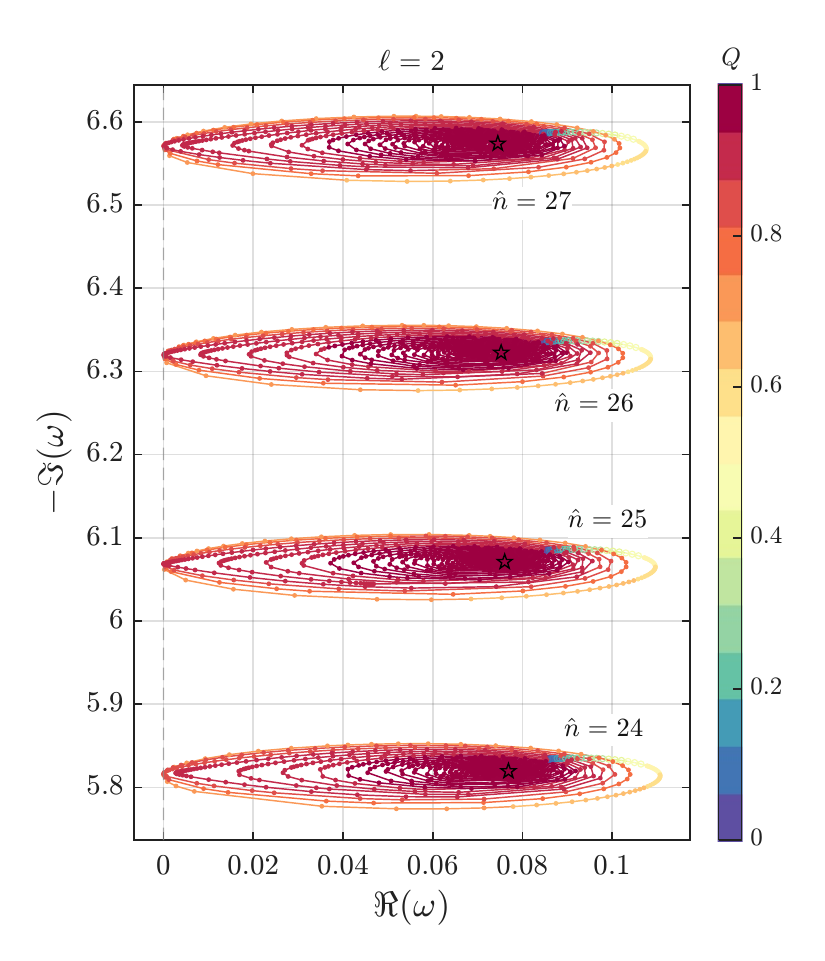}\hspace{.015\textwidth}\includegraphics[width=.49\textwidth]{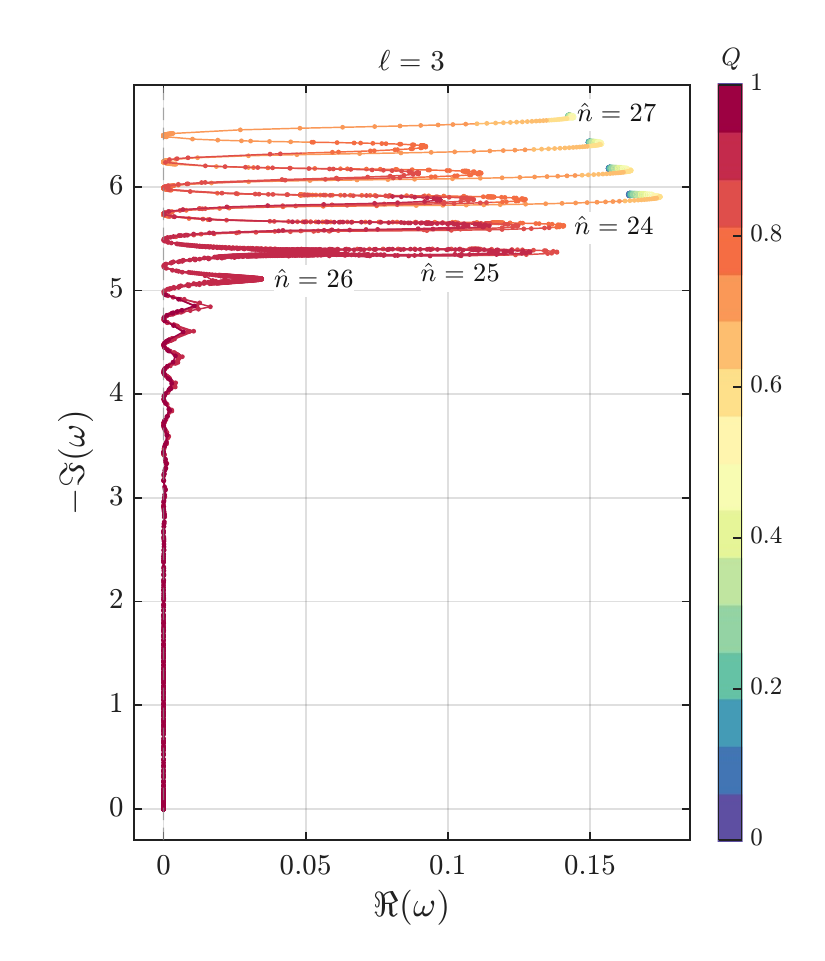}}\par\vspace{2pt}
\makebox[\textwidth][c]{\includegraphics[width=.49\textwidth]{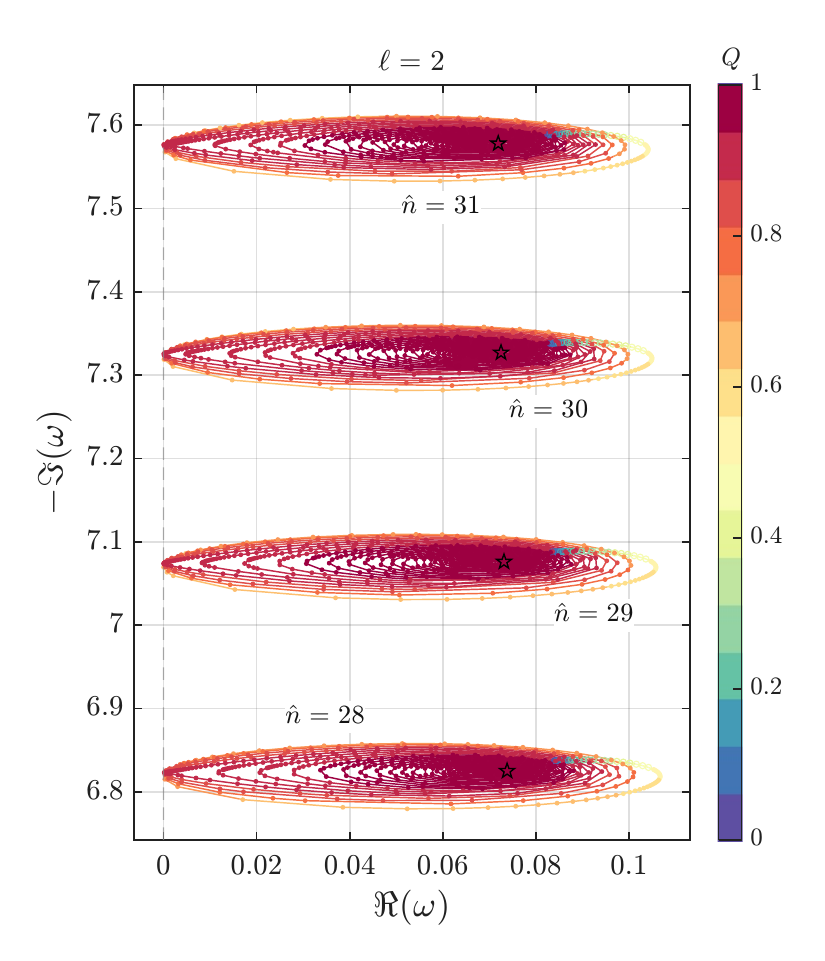}\hspace{.015\textwidth}\includegraphics[width=.49\textwidth]{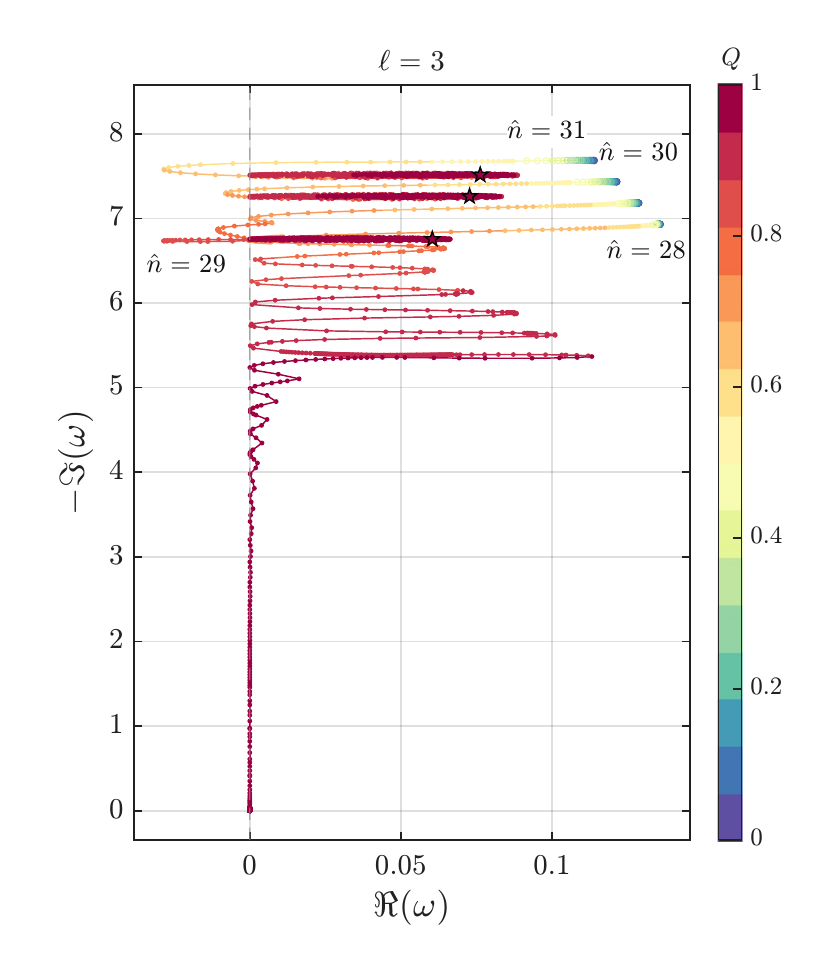}}
\caption{Overtones $\hat n=24$--$27$ (top) and $\hat n=28$--$31$ (bottom).}
\label{fig:atlas4}
\end{figure}

\begin{figure}[p]
\centering
\setlength{\abovecaptionskip}{3pt}
\makebox[\textwidth][c]{\includegraphics[width=.49\textwidth]{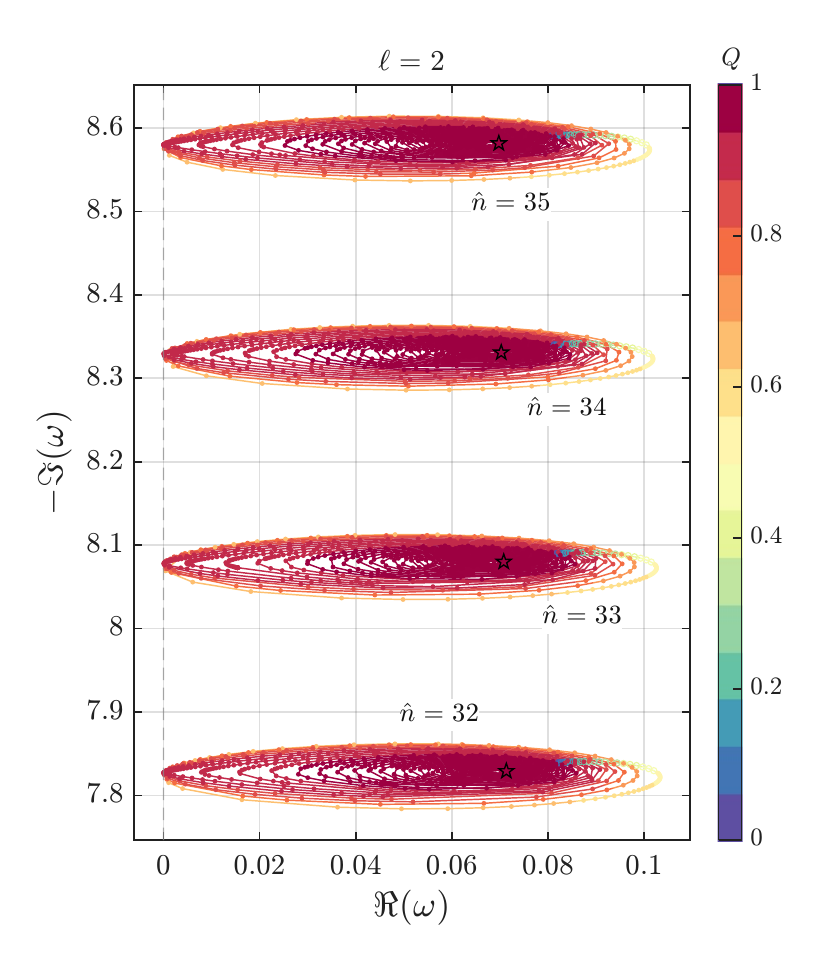}\hspace{.015\textwidth}\includegraphics[width=.49\textwidth]{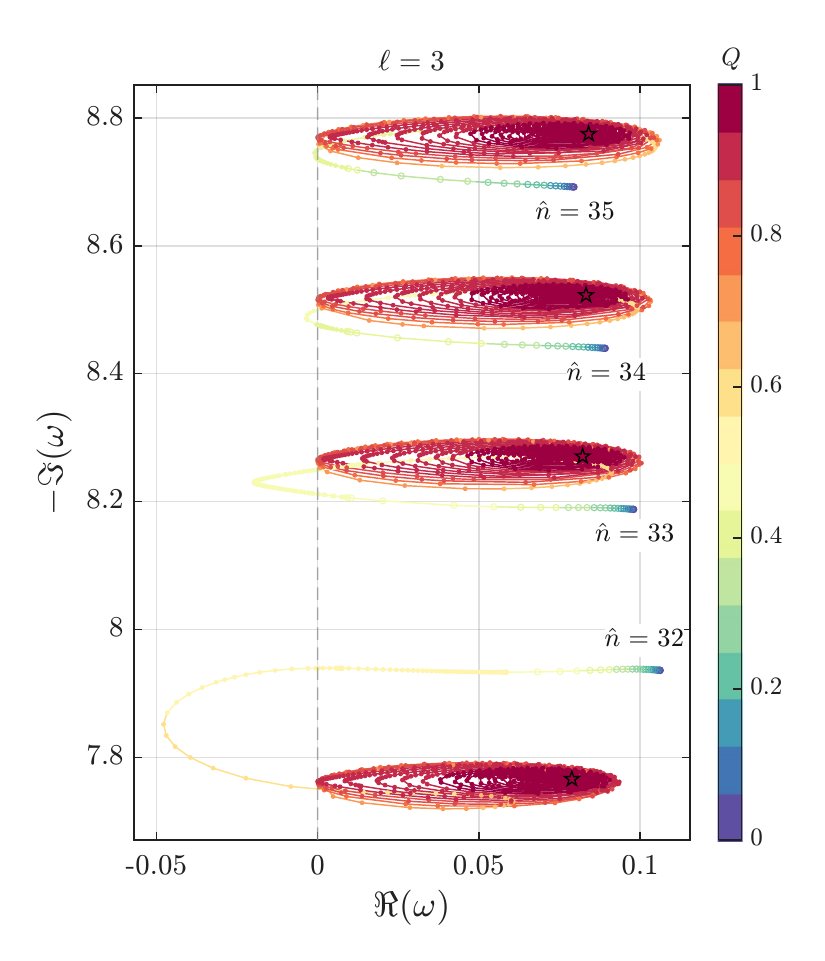}}\par\vspace{2pt}
\makebox[\textwidth][c]{\includegraphics[width=.49\textwidth]{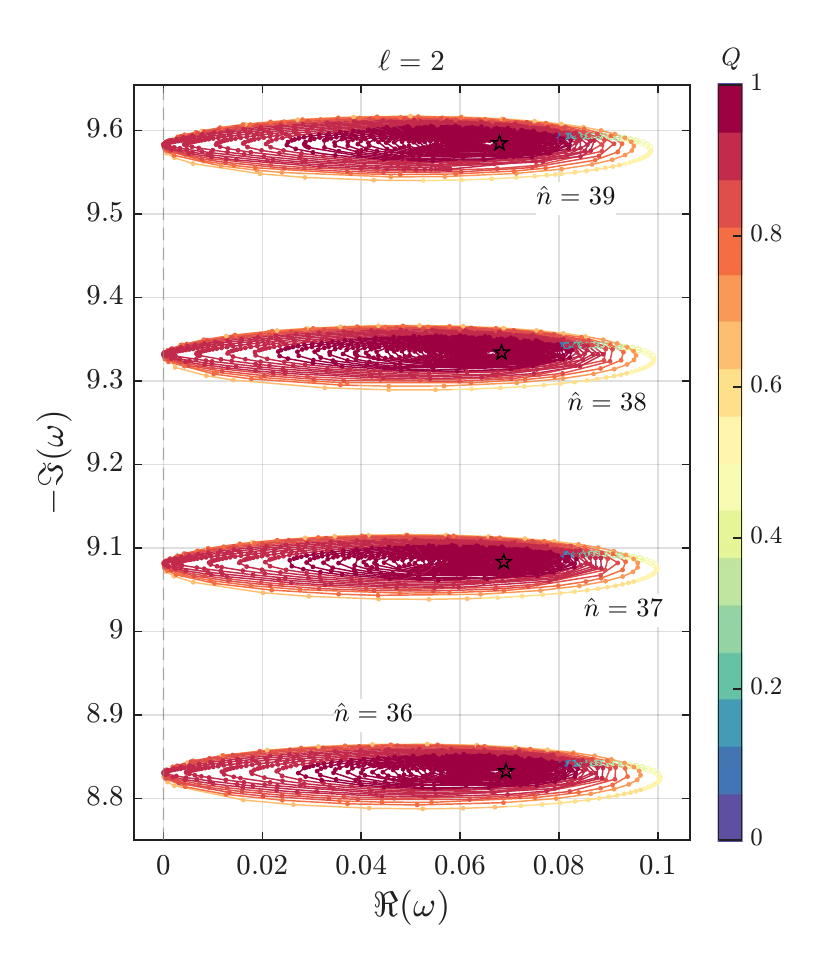}\hspace{.015\textwidth}\includegraphics[width=.49\textwidth]{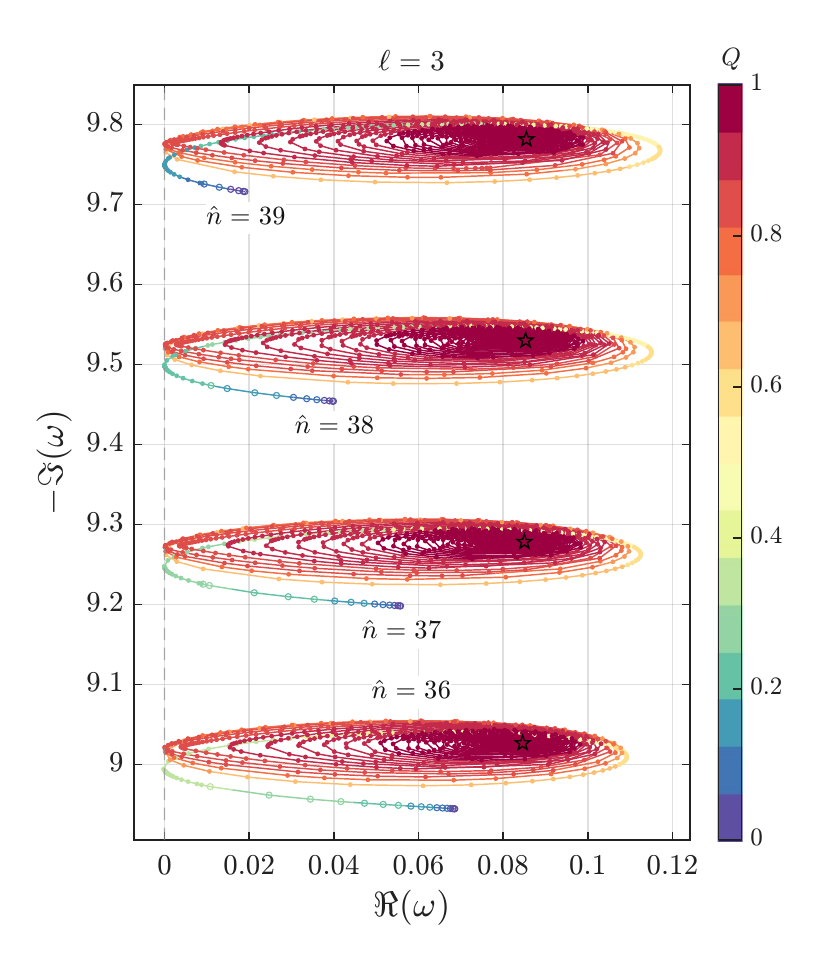}}
\caption{Overtones $\hat n=32$--$35$ (top) and $\hat n=36$--$39$ (bottom).}
\label{fig:representativecross}
\end{figure}

Figure~\ref{fig:bandstrict} shows the intermediate-damping spirals at $\ell=3$ and the three extremal frequencies in that band. For a comparison of the two angular orders, Figs.~\ref{fig:atlas1} and~\ref{fig:atlas2} cover $\hat n=0$--15, Figs.~\ref{fig:atlas3} and~\ref{fig:atlas4} cover $16$--31, and Figs.~\ref{fig:representativecross} and~\ref{fig:atlas5} cover $32$--41.

\begin{table}[tbp]
\caption{Gravitational QNMs on the unconventional sheet of extremal RN ($Q=1$). The amplitudes are evaluated at the unrounded frequencies.}
\label{tab:negativefinite}\centering\small
\begin{tabular}{crrrr}\toprule
& & & \multicolumn{2}{c}{$|B^{\rm inc}|$} \\
\cmidrule(lr){4-5}
$\ell$ & $\Re(\omega)$ & $\Im(\omega)$ & Unconventional sheet & Conventional sheet \\
\midrule
2 & $-0.009243460646$ & $-1.517870449784$ & $<4.0\times10^{-38}$ & $0.2680$ \\
3 & $-0.036729214653$ & $-6.032037006042$ & $<1.1\times10^{-43}$ & $0.5163$ \\
\bottomrule
\end{tabular}
\end{table}

\begin{table}[tbp]
\caption{Extremal limits of the gravitational overtones. Finite-damping frequencies are computed at $Q=1$. $\omega\to0$ denotes a family that approaches zero frequency as $Q\to1^-$.}
\label{tab:xrall}\centering\small
\begin{tabular}{crr}\toprule
$\hat n$ & $\ell=2$ & $\ell=3$ \\
\midrule
0 & $0.431340800668-0.083460315127\ii$ & $0.704304011534-0.085973373893\ii$ \\
1 & $0.404523553687-0.254984369586\ii$ & $0.688042245561-0.259923247954\ii$ \\
2 & $0.353401272686-0.441372603477\ii$ & $0.656241900434-0.440068867709\ii$ \\
3 & $0.287116330159-0.652454691130\ii$ & $0.610935787488-0.630911724607\ii$ \\
4 & $0.219944559830-0.888568320432\ii$ & $0.556324835439-0.836517999599\ii$ \\
5 & $0.156873316088-1.140220563874\ii$ & $0.498717775350-1.058639928985\ii$ \\
6 & $\omega\to0$ & $0.444347586579-1.295306872499\ii$ \\
7 & $0.079003520266-1.752910948774\ii$ & $0.396758062315-1.542088961964\ii$ \\
8 & $\omega\to0$ & $0.356542343074-1.794676037131\ii$ \\
9 & $0.089937490641-2.016759504137\ii$ & $0.322755346677-2.050131790682\ii$ \\
10 & $0.092405788104-2.275522936224\ii$ & $0.294089846666-2.306786399301\ii$ \\
11 & $0.092378259741-2.532193272014\ii$ & $0.269368761815-2.563772331634\ii$ \\
12 & $0.091416234374-2.787641644747\ii$ & $0.247653937503-2.820663389095\ii$ \\
13 & $0.090090601233-3.042256270985\ii$ & $0.228221948165-3.077263546954\ii$ \\
14 & $0.088643369289-3.296252834622\ii$ & $0.210511600423-3.333494575089\ii$ \\
15 & $0.087184871532-3.549767906790\ii$ & $0.194074543250-3.589337976500\ii$ \\
16 & $0.085766556136-3.802895431904\ii$ & $0.178534265234-3.844805219543\ii$ \\
17 & $0.084411634023-4.055703710771\ii$ & $0.163550712374-4.099922623808\ii$ \\
18 & $0.083129186100-4.308244372216\ii$ & $0.148784691358-4.354724006439\ii$ \\
19 & $0.081921080524-4.560557579614\ii$ & $0.133852365920-4.609247669128\ii$ \\
20 & $0.080785521053-4.812675274701\ii$ & $0.118249796520-4.863535976556\ii$ \\
21 & $0.079718924499-5.064623311583\ii$ & $\omega\to0$ \\
22 & $0.078716932365-5.316422921443\ii$ & $\omega\to0$ \\
23 & $0.077774958585-5.568091752272\ii$ & $0.101194447852-5.117635544134\ii$ \\
24 & $0.076888482454-5.819644627289\ii$ & $\omega\to0$ \\
25 & $0.076053199228-6.071094110689\ii$ & $\omega\to0$ \\
26 & $0.075265090557-6.322450937688\ii$ & $\omega\to0$ \\
27 & $0.074520449801-6.573724346691\ii$ & $\omega\to0$ \\
28 & $0.073815882347-6.824922339432\ii$ & $\omega\to0$ \\
29 & $0.073148292545-7.076051887203\ii$ & $0.060401011159-6.755113611557\ii$ \\
30 & $0.072514864013-7.327119096097\ii$ & $0.072744183193-7.261220311775\ii$ \\
31 & $0.071913037242-7.578129340703\ii$ & $0.076263617827-7.513867613427\ii$ \\
32 & $0.071340486697-7.829087373201\ii$ & $0.078829043806-7.766317101697\ii$ \\
33 & $0.070795098669-8.079997413104\ii$ & $0.082154924948-8.270746387150\ii$ \\
34 & $0.070274950507-8.330863221600\ii$ & $0.083221645991-8.522768035636\ii$ \\
35 & $0.069778291505-8.581688163550\ii$ & $0.084016865965-8.774681011897\ii$ \\
36 & $0.069303525544-8.832475259509\ii$ & $0.084601703045-9.026496584522\ii$ \\
37 & $0.068849195444-9.083227229624\ii$ & $0.085021188957-9.278224046609\ii$ \\
38 & $0.068413968929-9.333946530866\ii$ & $0.085309219350-9.529871216446\ii$ \\
39 & $0.067996626085-9.584635388777\ii$ & $0.085491766115-9.781444778715\ii$ \\
40 & $0.067596048174-9.835295824652\ii$ & $0.080731562330-8.018602060069\ii$ \\
41 & $0.067211207661-10.085929678923\ii$ & $0.085589031502-10.032950524811\ii$ \\
\bottomrule
\end{tabular}
\end{table}

\begin{table}[!htbp]
\caption{Subextremal frequencies at the largest sampled charge for each $\ell=2$ finite-damping family. The last column gives the distance to its $Q=1$ root.}
\label{tab:latestfinite2}\centering\small
\begin{tabular*}{.96\textwidth}{@{\extracolsep{\fill}}crrrr@{}}\toprule
$\hat n$ & $1-Q$ & $\Re(\omega)$ & $\Im(\omega)$ & $|\omega(Q)-\omega(1)|$ \\
\midrule
0 & $4.70562\times10^{-9}$ & $0.4313407997$ & $-0.0834603155$ & $1.05\times10^{-9}$ \\
1 & $1.28557\times10^{-8}$ & $0.4045235525$ & $-0.2549843722$ & $2.92\times10^{-9}$ \\
2 & $4.27745\times10^{-8}$ & $0.3534012751$ & $-0.4413726127$ & $9.54\times10^{-9}$ \\
3 & $1.05372\times10^{-7}$ & $0.2871163461$ & $-0.6524547057$ & $2.16\times10^{-8}$ \\
4 & $1.96518\times10^{-7}$ & $0.2199445941$ & $-0.8885683322$ & $3.62\times10^{-8}$ \\
5 & $6.64640\times10^{-7}$ & $0.1568734381$ & $-1.1402205722$ & $1.22\times10^{-7}$ \\
7 & $1.49653\times10^{-6}$ & $0.0790034921$ & $-1.7529108698$ & $8.38\times10^{-8}$ \\
9 & $4.33096\times10^{-6}$ & $0.0899375680$ & $-2.0167593615$ & $1.62\times10^{-7}$ \\
10 & $4.16031\times10^{-6}$ & $0.0924058949$ & $-2.2755228190$ & $1.59\times10^{-7}$ \\
11 & $4.16031\times10^{-6}$ & $0.0923783745$ & $-2.5321931629$ & $1.58\times10^{-7}$ \\
12 & $4.24478\times10^{-6}$ & $0.0914163515$ & $-2.7876415383$ & $1.58\times10^{-7}$ \\
13 & $4.33096\times10^{-6}$ & $0.0900907179$ & $-3.0422561659$ & $1.57\times10^{-7}$ \\
14 & $4.41890\times10^{-6}$ & $0.0886434843$ & $-3.2962527303$ & $1.55\times10^{-7}$ \\
15 & $4.60017\times10^{-6}$ & $0.0871849867$ & $-3.5497678009$ & $1.56\times10^{-7}$ \\
16 & $4.69357\times10^{-6}$ & $0.0857666690$ & $-3.8028953265$ & $1.54\times10^{-7}$ \\
17 & $4.88610\times10^{-6}$ & $0.0844117469$ & $-4.0557036035$ & $1.56\times10^{-7}$ \\
18 & $4.98531\times10^{-6}$ & $0.0831292968$ & $-4.3082442652$ & $1.54\times10^{-7}$ \\
19 & $5.08653\times10^{-6}$ & $0.0819211892$ & $-4.5605574727$ & $1.52\times10^{-7}$ \\
20 & $5.29518\times10^{-6}$ & $0.0807856301$ & $-4.8126751657$ & $1.54\times10^{-7}$ \\
21 & $5.40270\times10^{-6}$ & $0.0797190319$ & $-5.0646232026$ & $1.53\times10^{-7}$ \\
22 & $5.51239\times10^{-6}$ & $0.0787170382$ & $-5.3164228125$ & $1.52\times10^{-7}$ \\
23 & $5.73851\times10^{-6}$ & $0.0777750652$ & $-5.5680916410$ & $1.54\times10^{-7}$ \\
24 & $5.85503\times10^{-6}$ & $0.0768885879$ & $-5.8196445159$ & $1.53\times10^{-7}$ \\
25 & $6.09521\times10^{-6}$ & $0.0760533057$ & $-6.0710939968$ & $1.56\times10^{-7}$ \\
26 & $6.34523\times10^{-6}$ & $0.0752651981$ & $-6.3224508213$ & $1.59\times10^{-7}$ \\
27 & $6.60552\times10^{-6}$ & $0.0745205587$ & $-6.5737242276$ & $1.61\times10^{-7}$ \\
28 & $6.73964\times10^{-6}$ & $0.0738159904$ & $-6.8249222200$ & $1.61\times10^{-7}$ \\
29 & $1.00000\times10^{-5}$ & $0.0731484487$ & $-7.0760517129$ & $2.34\times10^{-7}$ \\
30 & $1.00000\times10^{-5}$ & $0.0725150162$ & $-7.3271189246$ & $2.29\times10^{-7}$ \\
31 & $1.00000\times10^{-5}$ & $0.0719131858$ & $-7.5781291719$ & $2.25\times10^{-7}$ \\
32 & $1.00000\times10^{-5}$ & $0.0713406317$ & $-7.8290872070$ & $2.21\times10^{-7}$ \\
33 & $1.00000\times10^{-5}$ & $0.0707952403$ & $-8.0799972494$ & $2.16\times10^{-7}$ \\
34 & $1.00000\times10^{-5}$ & $0.0702750890$ & $-8.3308630604$ & $2.13\times10^{-7}$ \\
35 & $1.00000\times10^{-5}$ & $0.0697784270$ & $-8.5816880046$ & $2.09\times10^{-7}$ \\
36 & $1.00000\times10^{-5}$ & $0.0693036582$ & $-8.8324751029$ & $2.05\times10^{-7}$ \\
37 & $1.00000\times10^{-5}$ & $0.0688493254$ & $-9.0832270751$ & $2.02\times10^{-7}$ \\
38 & $1.00000\times10^{-5}$ & $0.0684140963$ & $-9.3339463785$ & $1.99\times10^{-7}$ \\
39 & $1.00000\times10^{-5}$ & $0.0679967510$ & $-9.5846352384$ & $1.95\times10^{-7}$ \\
40 & $1.00000\times10^{-5}$ & $0.0675961708$ & $-9.8352956763$ & $1.92\times10^{-7}$ \\
41 & $1.00000\times10^{-5}$ & $0.0672113280$ & $-10.0859295325$ & $1.90\times10^{-7}$ \\
\bottomrule
\end{tabular*}
\end{table}

\begin{table}[p]
\caption{Subextremal frequencies at the largest sampled charge for each $\ell=3$ finite-damping family. The last column gives the distance to its $Q=1$ root.}
\label{tab:latestfinite3}\centering\small
\renewcommand{\arraystretch}{.92}
\begin{tabular*}{.96\textwidth}{@{\extracolsep{\fill}}crrrr@{}}\toprule
$\hat n$ & $1-Q$ & $\Re(\omega)$ & $\Im(\omega)$ & $|\omega(Q)-\omega(1)|$ \\
\midrule
0 & $8.29039\times10^{-9}$ & $0.7043040080$ & $-0.0859733748$ & $3.68\times10^{-9}$ \\
1 & $2.25697\times10^{-8}$ & $0.6880422381$ & $-0.2599232547$ & $1.00\times10^{-8}$ \\
2 & $4.24971\times10^{-8}$ & $0.6562418938$ & $-0.4400688853$ & $1.88\times10^{-8}$ \\
3 & $2.05957\times10^{-7}$ & $0.6109357939$ & $-0.6309118130$ & $8.86\times10^{-8}$ \\
4 & $2.91523\times10^{-7}$ & $0.5563248884$ & $-0.8365181051$ & $1.18\times10^{-7}$ \\
5 & $5.77447\times10^{-7}$ & $0.4987179276$ & $-1.0586400773$ & $2.13\times10^{-7}$ \\
6 & $1.30967\times10^{-6}$ & $0.4443479612$ & $-1.2953070805$ & $4.28\times10^{-7}$ \\
7 & $1.92033\times10^{-6}$ & $0.3967585926$ & $-1.5420891295$ & $5.56\times10^{-7}$ \\
8 & $2.60829\times10^{-6}$ & $0.3565430095$ & $-1.7946761433$ & $6.75\times10^{-7}$ \\
9 & $7.68842\times10^{-7}$ & $0.3227555269$ & $-2.0501317995$ & $1.80\times10^{-7}$ \\
10 & $9.78572\times10^{-7}$ & $0.2940900582$ & $-2.3067863924$ & $2.12\times10^{-7}$ \\
11 & $1.19643\times10^{-6}$ & $0.2693690029$ & $-2.5637723090$ & $2.42\times10^{-7}$ \\
12 & $1.43369\times10^{-6}$ & $0.2476542102$ & $-2.8206633508$ & $2.75\times10^{-7}$ \\
13 & $1.71800\times10^{-6}$ & $0.2282222607$ & $-3.0772634920$ & $3.17\times10^{-7}$ \\
14 & $1.99753\times10^{-6}$ & $0.2105119524$ & $-3.3334945039$ & $3.59\times10^{-7}$ \\
15 & $2.32256\times10^{-6}$ & $0.1940749451$ & $-3.5893378879$ & $4.12\times10^{-7}$ \\
16 & $2.67346\times10^{-6}$ & $0.1785347261$ & $-3.8448051127$ & $4.73\times10^{-7}$ \\
17 & $3.04660\times10^{-6}$ & $0.1635512442$ & $-4.0999224980$ & $5.46\times10^{-7}$ \\
18 & $3.41816\times10^{-6}$ & $0.1487853074$ & $-4.3547238616$ & $6.33\times10^{-7}$ \\
19 & $3.81966\times10^{-6}$ & $0.1338530943$ & $-4.6092475036$ & $7.47\times10^{-7}$ \\
20 & $4.26401\times10^{-6}$ & $0.1182506860$ & $-4.8635357875$ & $9.09\times10^{-7}$ \\
23 & $4.71721\times10^{-6}$ & $0.1011955794$ & $-5.1176353296$ & $1.15\times10^{-6}$ \\
29 & $6.67224\times10^{-6}$ & $0.0604002030$ & $-6.7551132666$ & $8.79\times10^{-7}$ \\
30 & $1.00000\times10^{-5}$ & $0.0727436755$ & $-7.2612198552$ & $6.83\times10^{-7}$ \\
31 & $1.00000\times10^{-5}$ & $0.0762632773$ & $-7.5138671753$ & $5.55\times10^{-7}$ \\
32 & $1.00000\times10^{-5}$ & $0.0788288187$ & $-7.7663166778$ & $4.80\times10^{-7}$ \\
33 & $1.00000\times10^{-5}$ & $0.0821548457$ & $-8.2707459845$ & $4.10\times10^{-7}$ \\
34 & $1.00000\times10^{-5}$ & $0.0832216147$ & $-8.5227676414$ & $3.96\times10^{-7}$ \\
35 & $1.00000\times10^{-5}$ & $0.0840168722$ & $-8.7746806250$ & $3.87\times10^{-7}$ \\
36 & $1.00000\times10^{-5}$ & $0.0846017391$ & $-9.0264962043$ & $3.82\times10^{-7}$ \\
37 & $1.00000\times10^{-5}$ & $0.0850212490$ & $-9.2782236725$ & $3.79\times10^{-7}$ \\
38 & $1.00000\times10^{-5}$ & $0.0853092990$ & $-9.5298708480$ & $3.77\times10^{-7}$ \\
39 & $1.00000\times10^{-5}$ & $0.0854918619$ & $-9.7814444156$ & $3.76\times10^{-7}$ \\
40 & $1.00000\times10^{-5}$ & $0.0807314207$ & $-8.0186016477$ & $4.36\times10^{-7}$ \\
41 & $1.00000\times10^{-5}$ & $0.0855891405$ & $-10.0329501666$ & $3.74\times10^{-7}$ \\
\bottomrule
\end{tabular*}
\par\medskip
\begingroup
\makeatletter\def\@captype{figure}\makeatother
\centering
\subfloat[Lower damping window]{\includegraphics[width=.477\textwidth,trim=0 9bp 0 8bp,clip]{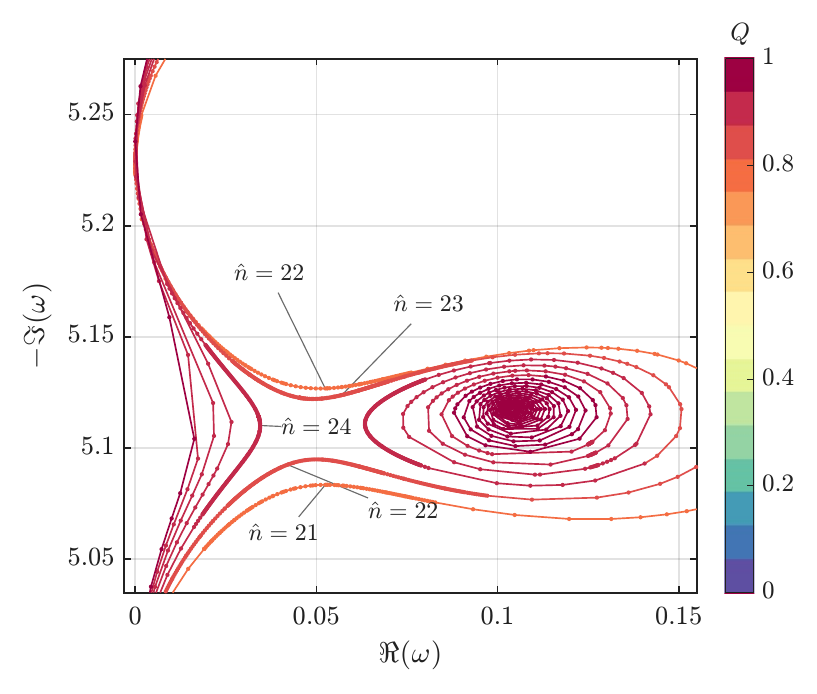}}\hfill
\subfloat[Higher damping window]{\includegraphics[width=.477\textwidth,trim=0 9bp 0 8bp,clip]{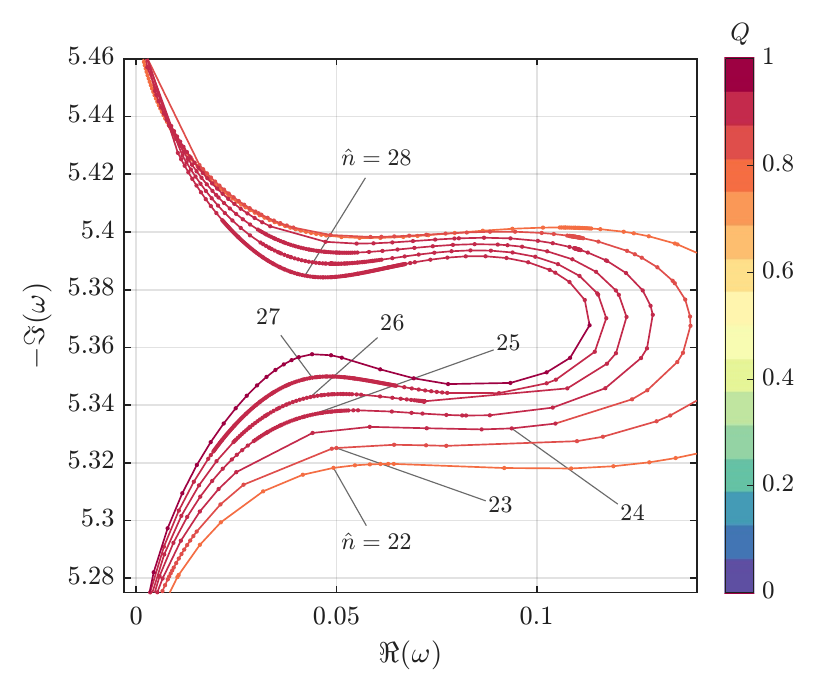}}
\caption{Lower- and higher-damping regions of the $\ell=3$ overtone encounters in Fig.~\ref{fig:pairapproach}. Color denotes $Q$.}
\label{fig:bandencounters}\label{fig:bandencountersmore}
\endgroup
\end{table}

\clearpage
\noindent\begin{minipage}{\textwidth}
\makeatletter\def\@captype{figure}\makeatother
\centering
\setlength{\abovecaptionskip}{3pt}
\makebox[\textwidth][c]{\includegraphics[width=.49\textwidth]{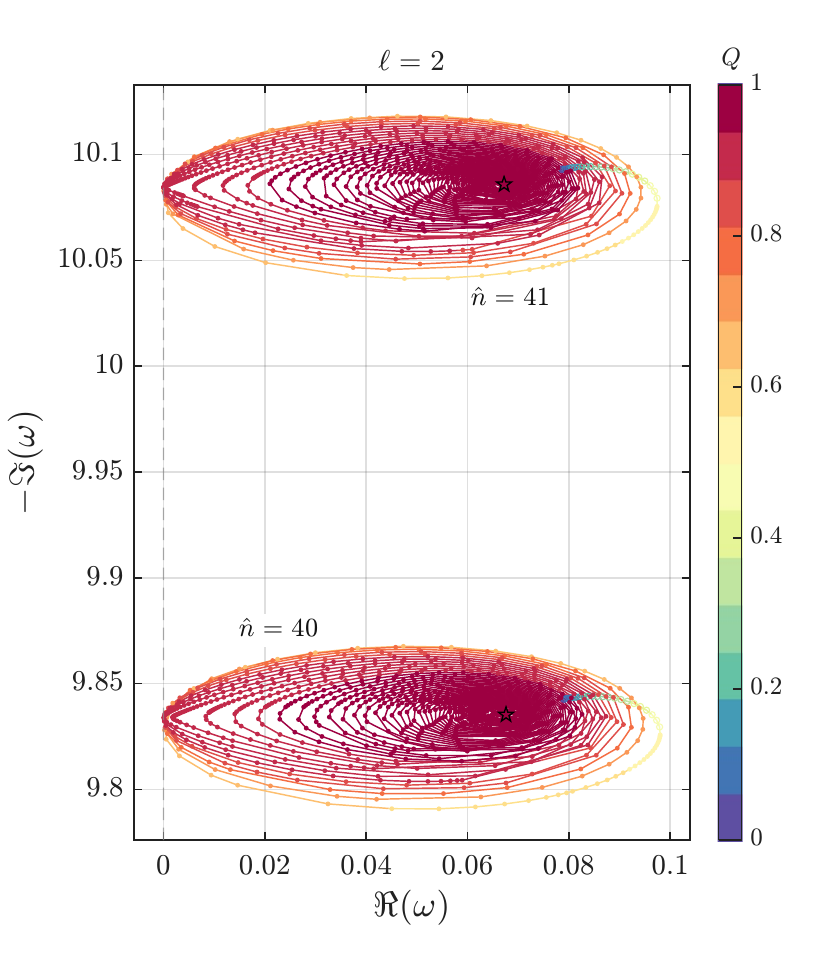}\hspace{.015\textwidth}\includegraphics[width=.49\textwidth]{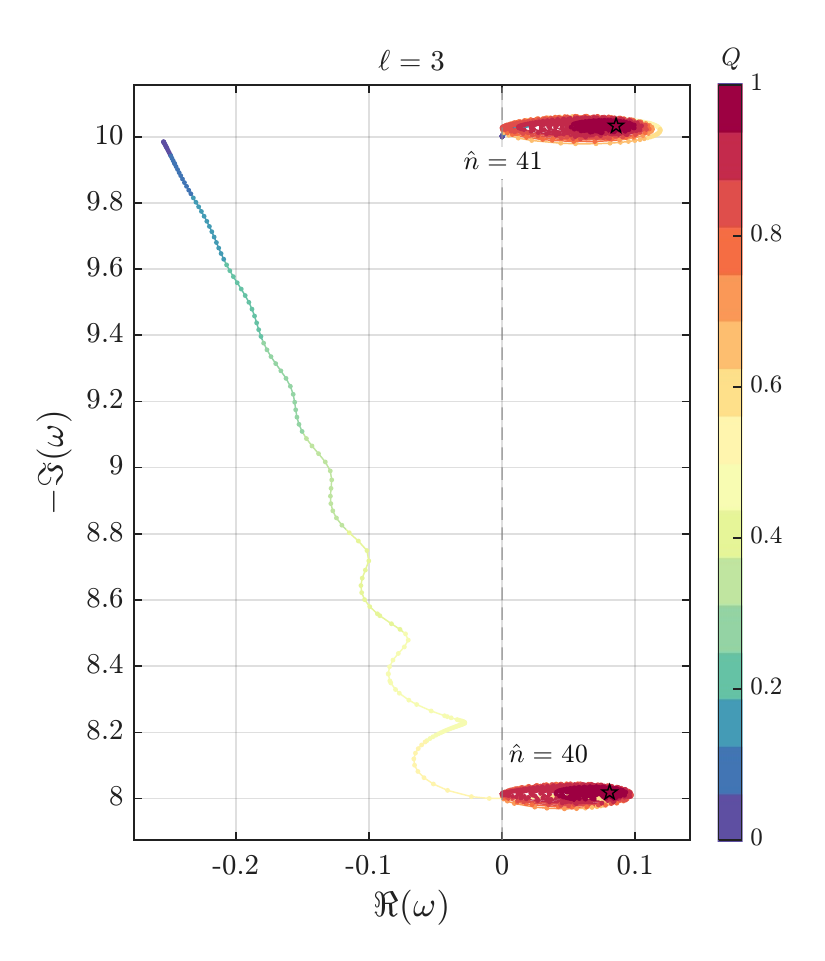}}
\caption{Overtones $\hat n=40$--$41$.}
\label{fig:atlas5}
\end{minipage}\par\medskip

\twocolumngrid
\bibliography{RN_refs}
\end{document}